\documentclass[fleqn,10pt]{article}
\usepackage[utf8]{inputenc}
\usepackage[T1]{fontenc}
\usepackage{lineno}

\usepackage{amssymb}
\usepackage{amsmath}
\usepackage{graphicx}
\usepackage{subcaption}
\usepackage{amssymb}
\usepackage{hyperref}
\usepackage{tabularray}
\usepackage{physics}
\usepackage[dvipsnames]{xcolor}
\usepackage{siunitx}
\usepackage{mathtools}
\usepackage{fullpage}
\usepackage{authblk}

\usepackage[colorinlistoftodos]{todonotes}
\presetkeys{todonotes}{inline}{}

\newcommand{\bt}[1]{\textbf{\textrm{#1}}}

\definecolor{mygreen}{RGB}{0,0,0}  

\title{Wallpaper Group-Based Mechanical Metamaterials: Dataset Including Mechanical Responses}

\author[1,2,*]{Fleur Hendriks}
\author[3,2]{Vlado Menkovski}
\author[4]{Martin Do\v{s}k\'{a}\v{r}}
\author[1,2]{Marc G.D. Geers}
\author[3,2]{Kevin Verbeek}
\author[1,2]{Ond\v{r}ej Roko\v{s}}

\affil[1]{Eindhoven University of Technology, Department of Mechanical Engineering, Eindhoven, 5600 MB, The Netherlands}
\affil[2]{Eindhoven AI Systems Institute (EAISI), Eindhoven, 5600 MB, The Netherlands}
\affil[3]{Eindhoven University of Technology, Department of Mathematics and Computer Science, Eindhoven, 5600 MB, The Netherlands}
\affil[4]{Czech Technical University in Prague, Faculty of Civil Engineering, Department of Mechanics, Prague 6, 166 29, Czech Republic}

\affil[*]{corresponding author(s): Fleur Hendriks (f.hendriks@tue.nl)}

\begin{document}

\flushbottom
\maketitle

\begin{abstract}
    Mechanical metamaterials often exhibit pattern transformations through instabilities, enabling applications in, e.g., soft robotics, sound reduction, and biomedicine. These transformations and their resulting mechanical properties are closely tied to the symmetries in these metamaterials' microstructures, which remain under-explored. Designing such materials is challenging due to the unbounded design space, and while machine learning offers promising tools, they require extensive training data. Here, we present a large dataset of 2D microstructures and their macroscopic mechanical responses in the hyperelastic, finite-strain regime, including buckling. The microstructures are generated using a novel method, which covers all 17 wallpaper symmetry groups and employs B\'ezier curves for a rich parametric space. Mechanical responses are obtained through finite element-based computational homogenization. The dataset includes 1,020 distinct geometries, each subjected to 12 loading trajectories, totaling 12,240 trajectories. Our dataset supports the development and benchmarking of surrogate models, facilitates the study of symmetry-property relationships, and enables investigations into symmetry-breaking during pattern transformations, potentially revealing emergent behavior in mechanical metamaterials.
\end{abstract}

\thispagestyle{empty}

\section*{Background \& Summary}

Mechanical metamaterials feature architected microstructures that are engineered to exhibit specific effective mechanical properties. To achieve this, designers often leverage buckling phenomena. Buckling causes sudden pattern transformations that result in an abrupt change in macroscopic behavior. Consequently, a metamaterial can switch between multiple distinct modes. These transformations enable tuning of acoustic metamaterials \cite{GuellIzard2020, Ning2021, Montgomery2021, Wu2021}, and can control the shape of a material, with applications in soft robotics \cite{Yang2015, Terryn2017, Kim2019} and biomedicine \cite{veerabagu2022}. These metamaterials can also exhibit a tunable negative Poisson's ratio \cite{Babaee2013} or negative compressibility \cite{Nicolaou2012}.

Here, we focus on flexible, porous 2D metamaterials revealing buckling-induced pattern transformations under mechanical loading \cite{Boyce2008, Overvelde2014} (neglecting other options such as pneumatics \cite{Yang2015}, or magnetic activation \cite{Kim2018}). These materials consist of a polymer base with holes of various shapes and sizes, typically arranged symmetrically to promote buckling. The geometry of these holes and their distribution controls mechanical properties such as auxeticity, anisotropy, and the bulk and shear moduli. Importantly, during buckling, microstructure symmetries are broken, making symmetry analysis essential to understand and predict deformation modes and their multiplicity \cite{Shim2013, Azulay2023, azulay2024predicting}. In 2D, these symmetries are described by the 17 wallpaper groups, which (in short IUCr notation) are \emph{p1}, \emph{p2}, \emph{pm}, \emph{pg}, \emph{cm}, \emph{pmm}, \emph{pmg}, \emph{pgg}, \emph{cmm}, \emph{p4}, \emph{p4m}, \emph{p4g}, \emph{p3}, \emph{p3m1}, \emph{p31m}, \emph{p6m} and \emph{p6} \cite{wallpapergroups, schattschneider1978plane}. The symmetries include translations, rotations (specified by a number that indicates the rotation order), reflections (indicated with \emph{m}), and glide reflections (indicated with \emph{g}). See Schattschneider \cite{schattschneider1978plane} for a detailed overview.

However, microstructural symmetry effects extend beyond buckling metamaterials, for example, chiral geometries are often exploited in auxetic materials \cite{Jin2019}. Note that although chiral metamaterials have proven their value in achieving auxeticity, chirality is neither necessary nor sufficient to achieve auxeticity (e.g., \cite{han2022lightweight}). In the same spirit, a mirror or glide mirror line ensures orthotropy of the effective behavior. The symmetries are also relevant when analyzing band gaps (Bragg diffraction), since they determine the shape of the irreducible Brillouin zone \cite{kittel2018introduction} and the location of the band-gap extremum within it \cite{Maurin2018}. They are also relevant for tailoring the dispersion of modes in acoustic metamaterials \cite{Moore2024} and for designing origami-based metamaterials \cite{Liu2021a, Liu2024}.

Currently, metamaterial datasets are scarce and limited in microstructure geometries, compromising the development and benchmarking of different design or analysis methods. Many datasets are limited to geometries based on periodic unit cells within the wallpaper groups \emph{pmm} \cite{Wang2020b, Chan2021, Bastek2023, Kollmann2020a} or \emph{p4m} \cite{Overvelde2014}, which inherently lack chirality and consistently result in orthotropic mechanical responses. Stochastic methods such as random distributions of ellipses, spinodal decomposition, Voronoi microstructures, or fractal noise \cite{Lyu123} only generate geometries within the group \emph{p1}. This misses highly symmetric geometries, such as \emph{p6m}, which can exhibit interesting multiplicities of buckling modes \cite{Ohno2001}. Moreover, these methods are often \textcolor{mygreen}{based on a limited number of building blocks with limited freedom (e.g. trusses)\cite{vangelatos2020regulating} or are} pixel-based \cite{Kollmann2020a, Lyu123, Wang2020b, Bastek2023}, with low resolution and often resulting in thin ligaments or disconnected structures, unless specifically constrained \cite{tyburec2025modular}. Therefore, sometimes the material volume fraction is bounded from below by 0.7 \cite{Chan2021}, leading to dense materials that do not buckle. Moreover, most datasets do not cover large deformations, as they only simulate one loading case in the reference configuration without a trajectory. Those that do \cite{Bastek2023, Zheng2023} are typically limited to uniaxial compression, which does not allow, e.g., analyzing of multiplicity of buckling modes as in honeycomb structures \cite{Ohno2001}.

The prime motivation behind our dataset of mechanical metamaterials with a large variety of geometries, is to facilitate the development of machine learning models, which are currently of significant interest \cite{Frankel2019, Wilt2020a, Pfaff2020, Vlassis2020,Pandey2021, Yang2021, Mianroodi2021, Mianroodi2022,Khorrami2023,Bastek2022,Bastek2023,Thomas2023, Karapiperis,Hendriks2025}. Currently, topology optimization is a staple for creating microstructures with specific mechanical responses \cite{Sigmund1994,Andreassen2014, Chen2018d,Thomsen2018, Wang2021, Xue2022b}. However, this method is computationally expensive, especially when large deformations and buckling are considered, and typically ends up in a local optimum. Moreover, it needs predefined domain boundaries of the unit cell (e.g. hexagonal, square, parallelogram). Machine learning-based methods hold promise as a faster alternative for simulation and generation of new designs \cite{Bastek2022, Zheng2023, Bastek2023}, and could rapidly generate a diverse set of potential solutions or provide probabilistic insights into the design space, increasing the chances of identifying a near-optimal or globally optimal configuration. However, they require a proper training dataset of microstructures with their corresponding mechanical responses.

This contribution addresses this need by providing a dataset that encompasses:
\begin{itemize}
    \item A large variety of geometries covering a broad design space of mechanical metamaterials, including all symmetry groups, based on a periodic graph `skeleton'.
    \item Well-structured, high-quality geometries: high resolution using B\'ezier curves, no too thin ligaments, fully connected (e.g., no laminates, no checkerboards), periodic, not too high volume fraction, and without sharp corners (to prevent stress concentrations).
    \item High-quality simulations covering large deformation responses, including buckling, to a variety of loading conditions.
\end{itemize}

\begin{figure}[htbp]
\centering
\includegraphics[width=\linewidth]{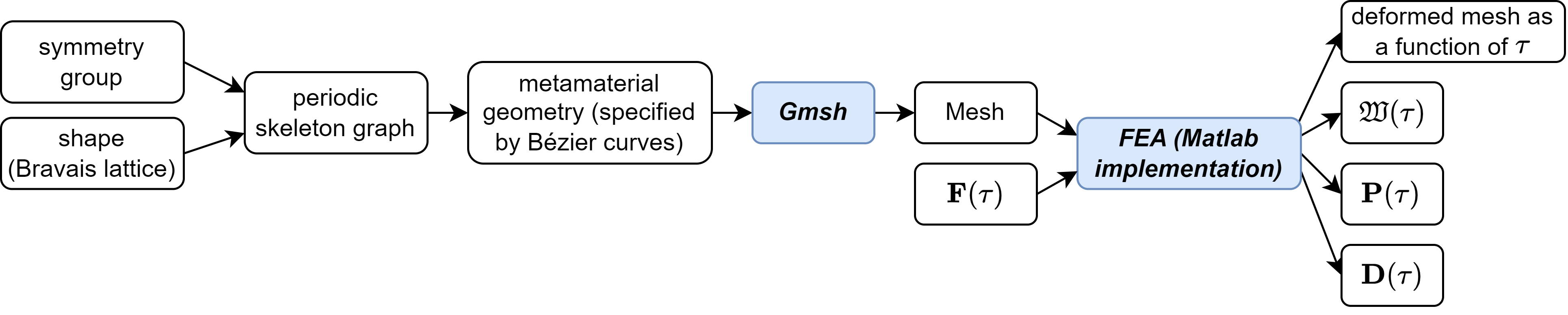}
\caption{Schematic overview of the creation process of the dataset.}
\label{fig:schematic_overview}
\end{figure}

Figure \ref{fig:schematic_overview} gives a schematic overview of the creation of the dataset, and Table \ref{tab:inoutputs} summarizes the input and output data. The dataset is available at \url{https://zenodo.org/records/15849549} \cite{datasetzenodo}; the Python code used to generate the geometries is available at \url{https://github.com/FHendriks11/wallpaper_microstructures}, and the MATLAB code used for the finite element simulations is available at \url{https://github.com/FHendriks11/mechmetamat_homogenization}.

\begin{table}[htbp]
\centering
\begin{tabular}{|l|l|l|}
\hline
\textbf{Input} & \textbf{Description} & \textbf{Shape, data type} \\
\hline
Geometry & Description of a mesh representing the geometry and of its periodicity & - \\
\hline
$\bt F$ & Deformation gradient tensor &  $(2,2)$ float\\
\hline
\textbf{Output} & \textbf{Description} & \textbf{Shape, data type} \\
\hline
$\vec{x}$ & Deformed configuration & $(N, 2)$ float \\
\hline
$\mathfrak{W}$ & Strain energy density & $(1,)$ float \\
\hline
$\bt P$ & First Piola-Kirchhoff stress tensor & $(2,2)$ float \\
\hline
$\bt D$ & Tangent stiffness tensor & $(2,2,2,2)$ float \\
\hline
\end{tabular}
\caption{\label{tab:inoutputs}The main inputs and outputs of the dataset.}
\end{table}

\section*{Methods}

In Figure \ref{fig:schematic_overview} it can be seen that our dataset was created in two steps: first, we generated the microstructure geometries, and then we computed their mechanical response for prescribed mascroscopic loading cases.
We will cover the steps of microstructure generation, sampling of the loading and simulation of the mechanical response in the sections below.

\subsection*{Generating new microstructures}
\begin{figure}[htbp]
\centering
\includegraphics[width=0.5\linewidth]{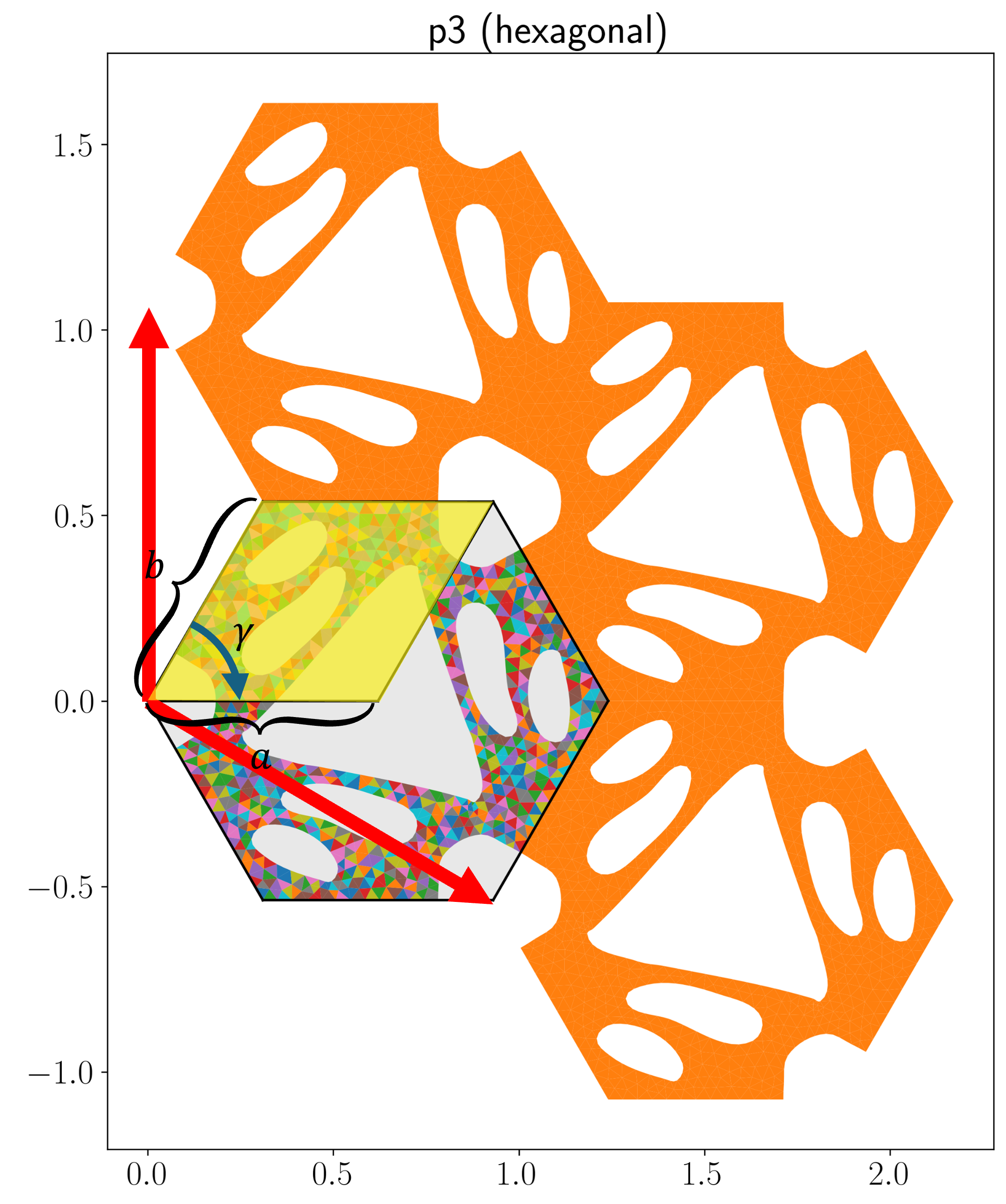}
\caption{Annotated example of a geometry with \emph{p3} symmetry. The fundamental domain is highlighted in yellow, the lattice vectors are shown as red arrows, and the side lengths $a$ and $b$ and the lower left corner $\gamma$ are indicated. The unit cell mesh is shown with a colorful mesh, and the rest of the entire RVE is orange.}
\label{fig:annotated_p3}
\end{figure}

\begin{figure}[htbp]
    \centering
    \includegraphics[scale=0.7]{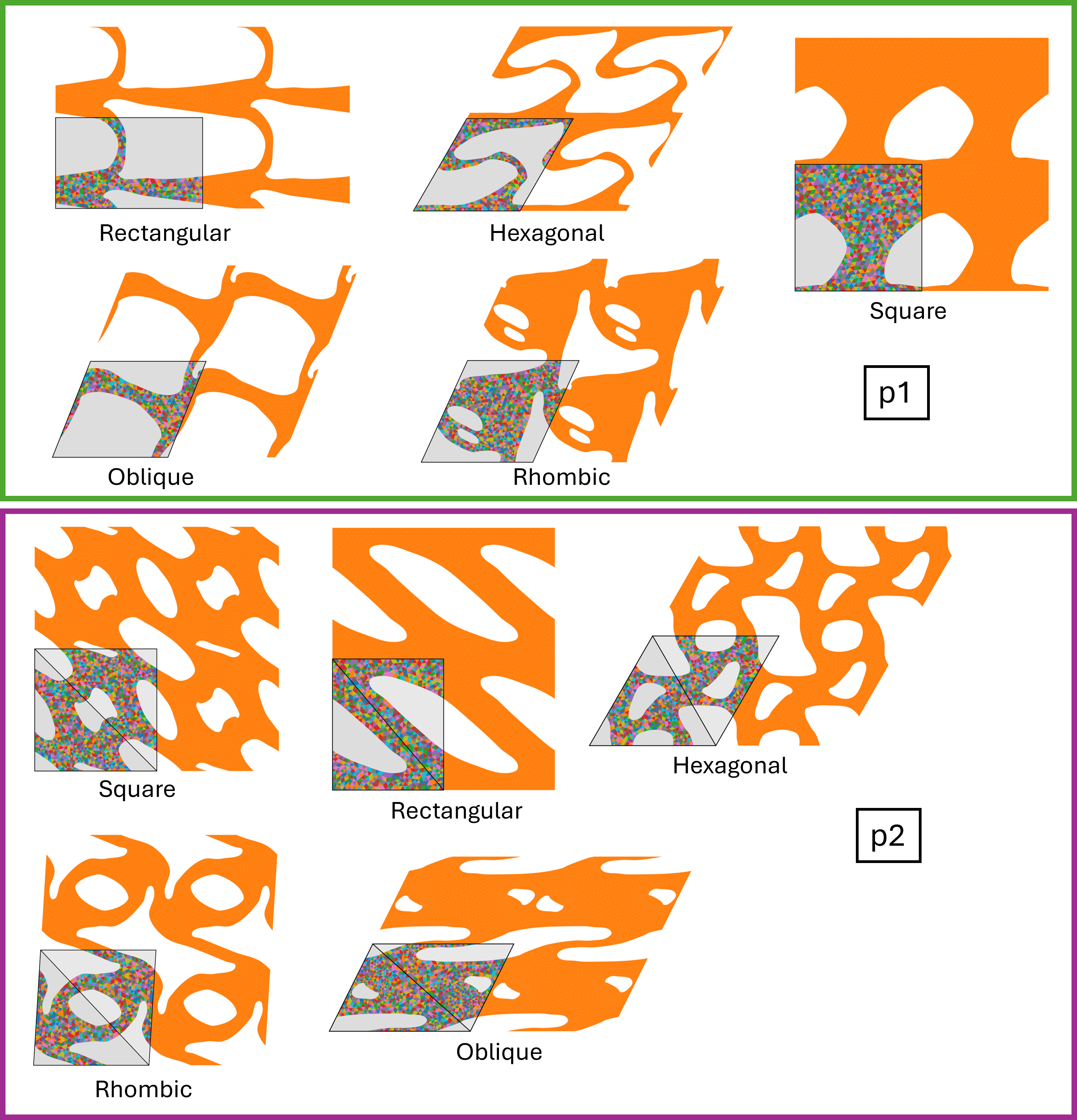}
    \caption{Examples of geometries corresponding to the wallpaper groups \emph{p1} and \emph{p2}. Each entire structure shown is one RVE, of which one unit cell is indicated by thick black lines and a light grey background. Within the unit cell, one fundamental domain is indicated, also with thick black lines and with a slightly darker grey background. Different Bravais lattices for the same wallpaper group are grouped together using colored boxes. The text in the white boxes indicates the the wallpaper group; the other text indicates the Bravais lattice.}
    \label{fig:examples_p1_p2}
\end{figure}

\begin{figure}[htbp]
    \centering
    \includegraphics[scale=0.7]{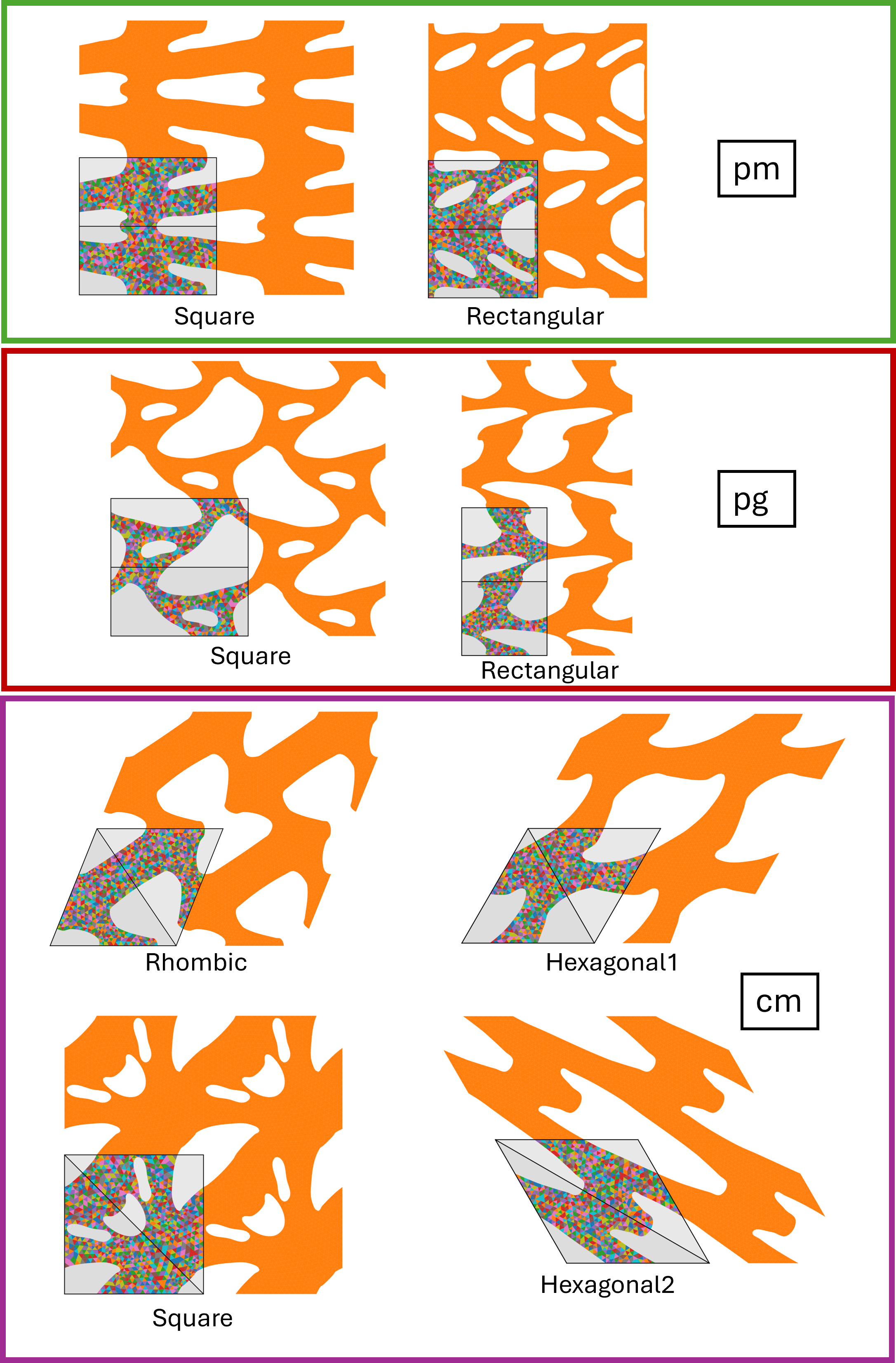}
    \caption{Examples of geometries corresponding to the wallpaper groups \emph{pm}, \emph{pg}, and \emph{cm}. Each entire structure shown is one RVE, of which one unit cell is indicated by thick black lines and a light grey background. Within the unit cell, one fundamental domain is indicated, also with thick black lines and with a slightly darker grey background. Different Bravais lattices for the same wallpaper group are grouped together using colored boxes. The text in the white boxes indicates the the wallpaper group; the other text indicates the Bravais lattice.}
    \label{fig:examples_pm_pg_cm}
\end{figure}

\begin{figure}[htbp]
    \centering
    \includegraphics[scale=0.7]{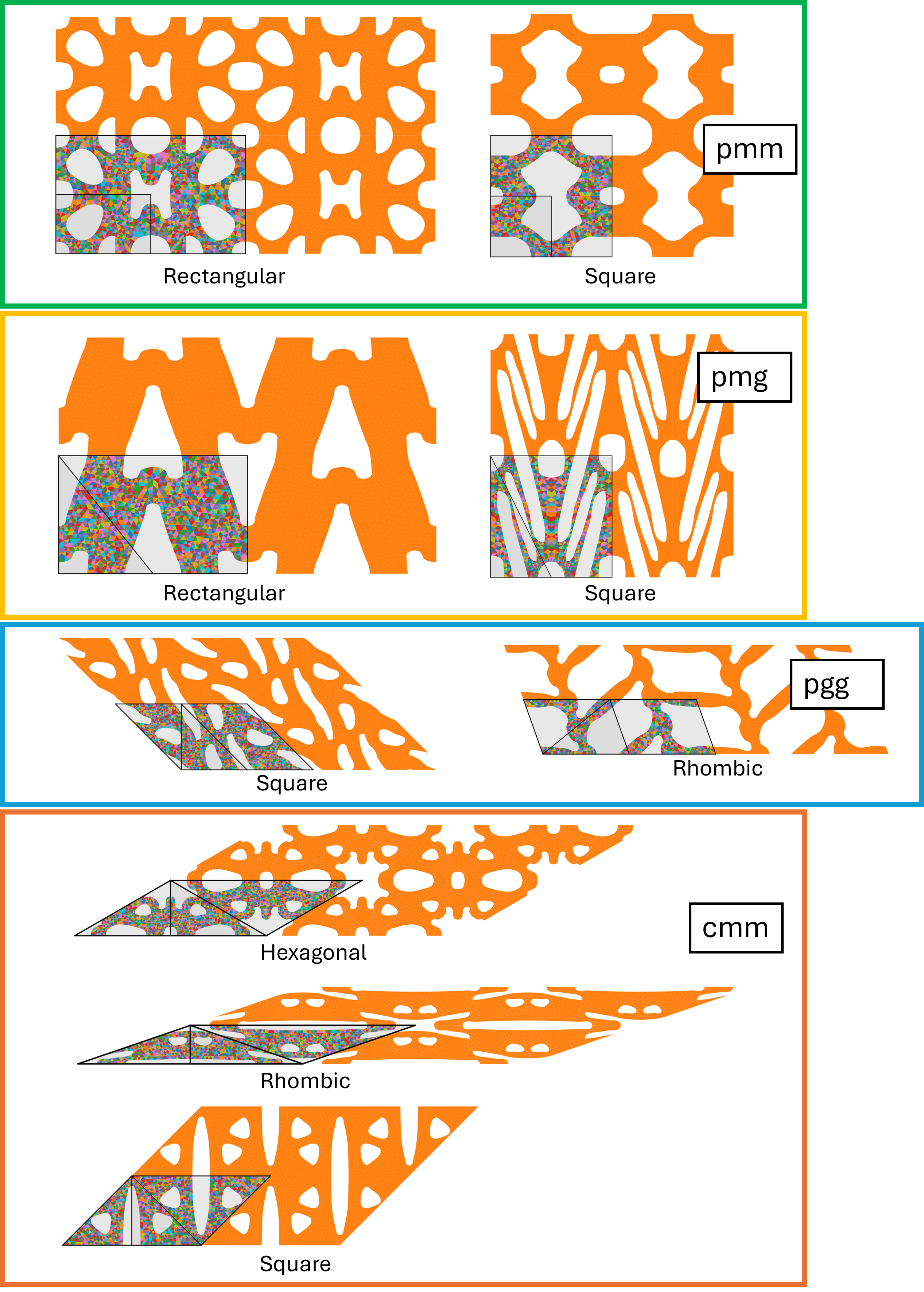}
    \caption{Examples of geometries corresponding to the wallpaper groups \emph{pmm}, \emph{pmg}, \emph{pgg}, and \emph{cmm}. Each entire structure shown is one RVE, of which one unit cell is indicated by thick black lines and a light grey background. Within the unit cell, one fundamental domain is indicated, also with thick black lines and with a slightly darker grey background. Different Bravais lattices for the same wallpaper group are grouped together using colored boxes. The text in the white boxes indicates the the wallpaper group; the other text indicates the Bravais lattice.}
    \label{fig:examples_pmm_pmg_pgg_cmm}
\end{figure}

\begin{figure}[htbp]
    \centering
    \includegraphics[scale=0.7]{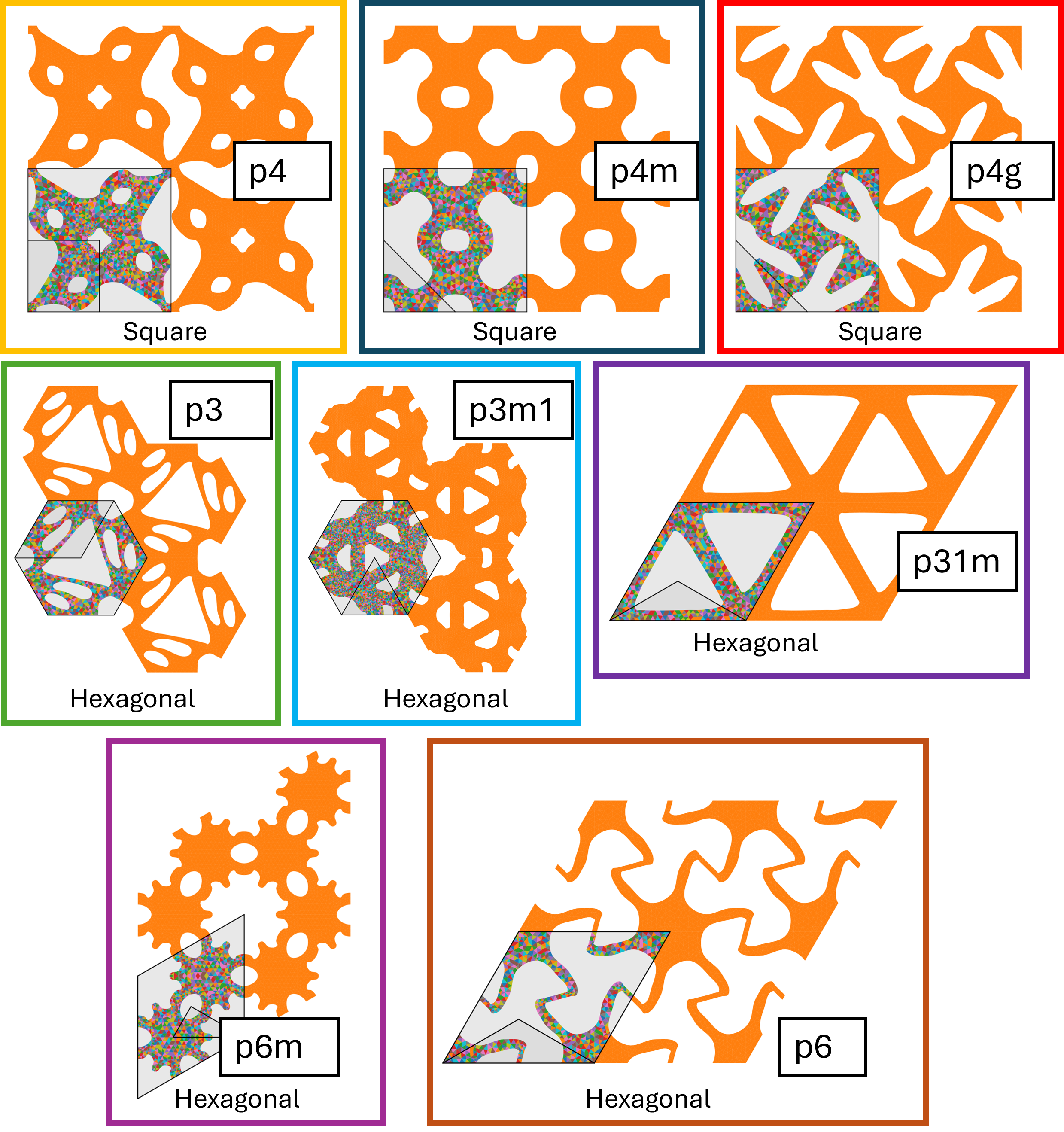}
    \caption{Examples of geometries corresponding to the wallpaper groups \emph{p4}, \emph{p4m}, \emph{p4g}, \emph{p3}, \emph{p3m1}, \emph{p31m}, \emph{p6m}, and \emph{p6}. Each entire structure shown is one RVE, of which one unit cell is indicated by thick black lines and a light grey background. Within the unit cell, one fundamental domain is indicated, also with thick black lines and with a slightly darker grey background. The text in the white boxes indicates the the wallpaper group; the other text indicates the Bravais lattice.}
    \label{fig:examples_p4_p4m_p4g_p3_p3m1_p31m_p6m_p6}
\end{figure}

All 2D periodic structures belong to one of 17 wallpaper groups \cite{wallpapergroups, schattschneider1978plane}, which categorize all possible ways to tile a 2D plane based on their symmetries, which include translations, rotations, reflections, and glide reflections. Each group represents a unique combination of these symmetries. In this framework, the unit cell is the smallest repeating unit that can create a periodic structure using only translations. The unit cell further consists of one or more copies of the fundamental domain, which tiles the space using not just translations, but also reflections, rotations, and/or glide reflections. The lattice vectors that define the unit cell can correspond to any of the 5 Bravais lattices in 2D. The Bravais lattices characterize the symmetries of a lattice, generated only by translating a point along the lattice vectors. In 2D, there are 5 Bravais lattice types: square, rectangular, hexagonal, centered rectangular (rhombic), and oblique, although not all Bravais lattice types are possible for all wallpaper groups. See Figure \ref{fig:annotated_p3} for an example showing the relevant geometric quantities, and Figure \ref{fig:examples_p1_p2},
\ref{fig:examples_pm_pg_cm}, \ref{fig:examples_pmm_pmg_pgg_cmm} and \ref{fig:examples_p4_p4m_p4g_p3_p3m1_p31m_p6m_p6} for examples of each wallpaper group and its possible Bravais lattices. Note that for the wallpapergroup \emph{cm}, there are two options to use a hexagonal Bravais lattice: one with the mirror line bisecting the unit cell along the short diagonal and the other along the long diagonal. We denote these two options `hexagonal1' and `hexagonal2', respectively.

\begin{figure}[htbp]
\centering
\includegraphics[width=\linewidth]{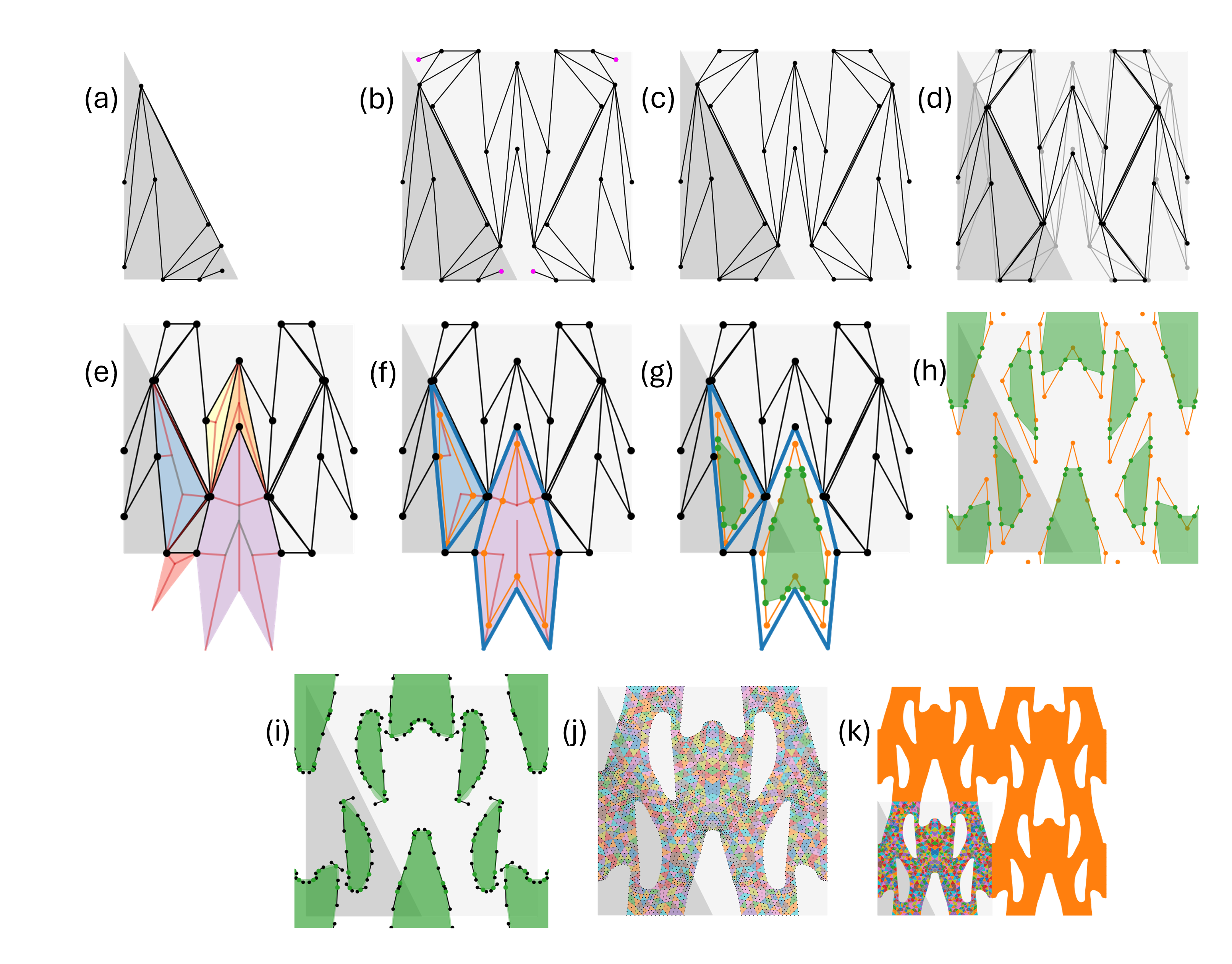}
\caption{The various steps in the generation of the microstructures. (a) Skeleton graph on the fundamental domain, (b) fundamental domain tiled into a unit cell, with leafs identified in magenta, (c) leafs removed, (d) nodes shifted to mitigate sharp angles, (e) unique faces (initial face polygons) with their straight skeleton, (f) secondary face polygon in orange, (g) tertiary face polygons approximating hole shape in green, (h) polygons copied around, (i) beziergons, (j) mesh of the unit cell, (k) final RVE.}
\label{fig:material_generation}
\end{figure}

In our method, a new geometry was generated using the following steps, shown in Figure \ref{fig:material_generation}:
\begin{itemize}
    \item Choose a wallpaper group and a Bravais lattice type.
    \item Generate the parallelogram or triangle that represents the shape of the fundamental domain.
    \item Generate a random periodic graph: the `skeleton graph'. This geometric graph should be planar and connected, and respect the chosen wallpaper group.
    \item Tile this graph into a graph covering the unit cell.
    \item Assume at first that all graph faces are filled with bulk material. Then create a hole on each face of the graph; each hole is described by a b\'eziergon, i.e., a closed path composed of B\'ezier curves.
    \item Discretize only one fundamental domain, respecting compatibility requirements, then tile this to obtain the unit cell discretization.
    \item Validate the obtained geometry throughout (e.g., check if the volume fraction is not too high, and that ligaments are not too thin) and discard invalid geometries.
\end{itemize}

This method of generating new microstructures is based on a periodic graph `skeleton', which roughly defines the connectivity of the microstructure.
This graph skeleton also offers a versatile backbone that can be used by machine learning methods such as graph neural networks \cite{Hendriks2025}. This method overcomes the limitations of previous approaches mentioned in the previous section by generating microstructures for any of the 17 wallpaper groups, with varying ligament thicknesses and hole shapes. These 2D materials can be turned into 3D materials by extruding them along the out-of-plane direction.

Other works have used periodic graphs to generate metamaterials, but these are limited to simple building blocks, like trusses, shells or cubes \cite{Bastek2022, Zheng2023, Abu-Mualla2024, KarimiMahabadi2025}.
In contrast, our geometries are specified by Bézier curves, making them smooth and highly resolved, allowing any arbitrary shape. The method also ensures that the microstructures are connected and periodic, and that the ligaments are unlikely to be thinner than a specified threshold (and the cases that do have too thin ligaments are rare and easily filtered out). The fact that other works have used graph-based approaches to generate 3D metamaterials, albeit in simpler forms, suggests that our method could be extended to 3D as well.

The microstructures were generated using Python code which is available at \url{https://github.com/FHendriks11/wallpaper_microstructures}.

We discuss the microstructure generation steps in order below.

\subsubsection*{Generating the fundamental domain shape}
Depending on the chosen wallpaper group we used either a parallelogram or a triangle as the fundamental domain shape.
Without loss of generality, we fixed the orientation of the fundamental domain by aligning its bottom edge along the $x$-axis, and then set the length of this edge to 1. We then chose the length of the left side $b$ (which is not necessarily vertical), as well as the lower-left corner angle $\gamma$. Depending on the wallpaper group and the Bravais lattice, $b$ and/or $\gamma$ could be fixed already, e.g., for a square Bravais lattice, $b$ was also 1 and $\gamma$ was 90 degrees. If they were not fixed, we picked them from a uniform range of values, specified for each combination of wallpaper group and Bravais lattice. Different values of $b$ and $\gamma$ lead to different geometries and therefore different material properties. See Figure \ref{fig:annotated_p3}, where $a$, $b$ and $\gamma$ are indicated for the wallpaper group \emph{p3}.
We then immediately rescaled $a$ and $b$ such that the fundamental domain has an area of $1/\textrm{(number of fundamentals domains in the unit cell)}$, to ensure the unit cell would have unit area.

\subsubsection*{Periodic skeleton graph generation}
For our skeleton graph, we desire a random geometric graph that allows for a lot of freedom in the shape of the faces (unlike, e.g., a Voronoi graph, which can only have convex faces), but still respect the right symmetries. To this end, we started with the fundamental domain, see Figure \ref{fig:material_generation}(a), and first randomly distributed points on its boundary. We made sure each edge of the boundary contained at least one point (a point on a vertex of the polygon involving that edge also counts). Next, we randomly generated points inside it.
Edges of the skeleton graph were also created stochastically: we randomly picked two points and checked whether an edge connecting them intersected already defined edges. If not, we added the new edge to the graph. We repeated this process until all points were connected to each other.
Checking the connectivity of a periodic graph is not as simple as checking connectivity of the fundamental domain graph; we explain our connectivity check in Appendix \ref{sec:appconn_check}.
By rotating, reflecting and translating the graph defined on the fundamental domain, we were able to create the skeleton graph of the unit cell, see Figure \ref{fig:material_generation}(b).
We further removed leafs (nodes of degree 1), and shifted some of the nodes of the graph to attempt to reduce the number of really sharp angles ($< \pi/7$), see Figures \ref{fig:material_generation}(c) and (d), respectively.

\subsubsection*{Creating the holes}
To distribute holes in a microstructure, we treated each face of the graph as a potential location for a hole. To this end, we followed this procedure, which we illustrate in Figure \ref{fig:material_generation}:
\begin{itemize}
    \item \emph{Construct primary polygons}. We first identified all the unique faces of the unit cell skeleton graph, i.e. those that are not transformed copies of each other, see Figure \ref{fig:material_generation}(e). This ensured that the transformed copies got the same hole shape. We call the shapes of the graph faces the primary polygons.
    \item \emph{Randomly select faces for hole creation} For each face, we randomly decided to either attempt to create a hole in it (with probability 98\%), or fill it with bulk material (probability 2\%).
    \item \emph{Construct secondary polygons: approximate the thickness of the ligaments (orange lines in Figure \ref{fig:material_generation}(f))}. To approximately determine the thickness of the ligaments, we created a secondary polygon on each graph face marked for hole generation, outlined with orange lines in Figure \ref{fig:material_generation}(f). To this end, we picked a point on each bisector of the corners of the primary polygon. To ensure that the secondary polygon did not self-overlap, we only allowed picking points on a certain segment of the bisector. To determine this segment, we used the straight skeleton, which is a well-known method of representing a polygon in computational geometry, for which efficient algorithms exist to compute it. This straight skeleton represents a continuous shrinking inwards of the polygon edges. We generated the straight skeleton of each face using the scikit-geometry interface to the CGAL library \cite{cgal:c-sspo2-24b}, see Figure \ref{fig:material_generation}(e).  We used the segments of the bisectors that are part of the straight skeleton (the red lines in Figure \ref{fig:material_generation}(e) and (f)) as our range to pick the new points, which ensured that the new polygon did not overlap itself. For each new point, we made sure it had a minimum distance $l_{min}$ to the corresponding vertex of the face. This minimum distance was calculated as
\begin{equation}
    d_{min} =
    \begin{cases}
        \texttt{MIN\_THICKNESS}/\sin\left(\frac{\theta^\text{prim}}{2}\right) & \quad \textrm{if the polygon is convex ($0 < \theta^\text{prim} \leq \pi$) at this corner} \\
        \texttt{MIN\_THICKNESS}/\cos\left(\frac{\theta^\text{prim}-\pi}{2}\right) & \quad \textrm{if the polygon is concave ($\theta^\text{prim} > \pi$) at this corner}
    \end{cases}
\end{equation}
where $\texttt{MIN\_THICKNESS}$ is the minimum thickness to add and $\theta^\text{prim}$ is the angle of the primary polygon corner. This ensured that the ligament thickness is at least $2\texttt{MIN\_THICKNESS}$, because the neighboring face can also add the same thickness. In other words, \verb|MIN_THICKNESS| is not the minimum thickness of the ligament itself, but of the ligament inside one face.
The maximum distance $d_{max}$ was \verb|MAX_REL_THICKNESS|$=0.7$ times the length of the bisector. If no point satisfies the minimum and maximum distance because $d_{min} > d_{max}$, then no hole was added to the face and the face was filled with bulk material.
We also checked if this polygon has an area larger than the parameter \verb|MIN_AREA|. If the area was smaller, the face was, again, filled with bulk material.
    \item \emph{Construct tertiary polygons: add material in the corners of the secondary polygon (green lines in Figure \ref{fig:material_generation}(g))}. So far, all the polygons had very sharp corners where ligaments come together. Therefore, we created a tertiary polygon inside the secondary one, which added material in the corners. This tertiary polygon is shown in green in Figure \ref{fig:material_generation}(g). We created it by randomly choosing two points on each edge of the secondary polygon which were (1) a minimum distance away from the corners, and (2) a minimum distance apart from each other. The minimum distance (1) was computed
\begin{equation}
    \textrm{Minimum distance} = \max\left(
            \frac{
                \text{\texttt{MIN\_RADIUS}}
                }{
                    \tan\left(
                        \theta^\text{sec}/2
                        \right)
                },
            \verb|MIN_D_REL| \times \textrm{length of edge}
        \right)
\end{equation}
where \verb|MIN_RADIUS| and \verb|MIN_D_REL| are parameters, which are the minimum radius of curvature, and a minimum distance relative to the length of the edge. $\theta^\text{sec}$ is the angle of the secondary polygon corner, in the interval $[0, \pi]$ (i.e., no distinction between convex and concave corners).
The minimum distance apart from each other (2) was computed as:
\begin{equation}
    \textrm{Minimum separation} = \max\left(\texttt{MIN\_SEP\_ABS}, \texttt{MIN\_SEP\_REL} \times \textrm{length of edge}\right)
\end{equation}
where \verb|MIN_SEP_ABS| and \verb|MIN_SEP_REL| are parameters, denoting the minimum absolute and relative distance between the two points on the edge, respectively. In creating the secondary and tertiary polygons, we respected the symmetry of the holes, by using the same randomly generated numbers for corresponding corners and edges.
    \item \emph{Replicate hole shapes across identical faces (Figure \ref{fig:material_generation}(h))}. After copying the tertiary polygons to all equivalent faces of the unit cell skeleton graph.
    \item \emph{Create beziergon to define the final hole shape (Figure \ref{fig:material_generation}(i))}. A closed composite cubic B\'ezier curve (a so-called beziergon) was drawn through all vertices of each tertiary polygon, rendering a smooth, continuous path defining the hole shape, see Figure \ref{fig:material_generation}(i). The corners of the tertiary polygon were used as the start and end points of the B\'ezier segments. Per segment, two extra control points were generated, which defined the direction of the tangents to the Bézier curve segment.
These control points were chosen such that (1) control points of two neighbouring curves around the joint between these curves are collinear with this joint (see for example control points $T_1$ and $T_2$ and the joint at point B in Figure \ref{fig:tangentpoints}; $T_1$, $T_2$ and B are collinear), (2) the line connecting them is perpendicular to the bisector of the corner of the tertiary polygon (see the line $T_1T_2$ and the red bisector in Figure \ref{fig:tangentpoints}) and (3) the distance between the joint and the new control points is equal to $1/3$ times the distance between the end points of the B\'ezier segment (e.g., $|T_1 B| = |AB|/3$, $|BT_2|=|BC|/3$ in Figure \ref{fig:tangentpoints}).
    \item \emph{Filter out invalid geometries.} The cases resulting in a too high volume fraction or unacceptable geometries were excluded from the dataset. See the validation section for details.
\end{itemize}

\begin{figure}[htbp]
\centering
\includegraphics[width=0.5\linewidth]{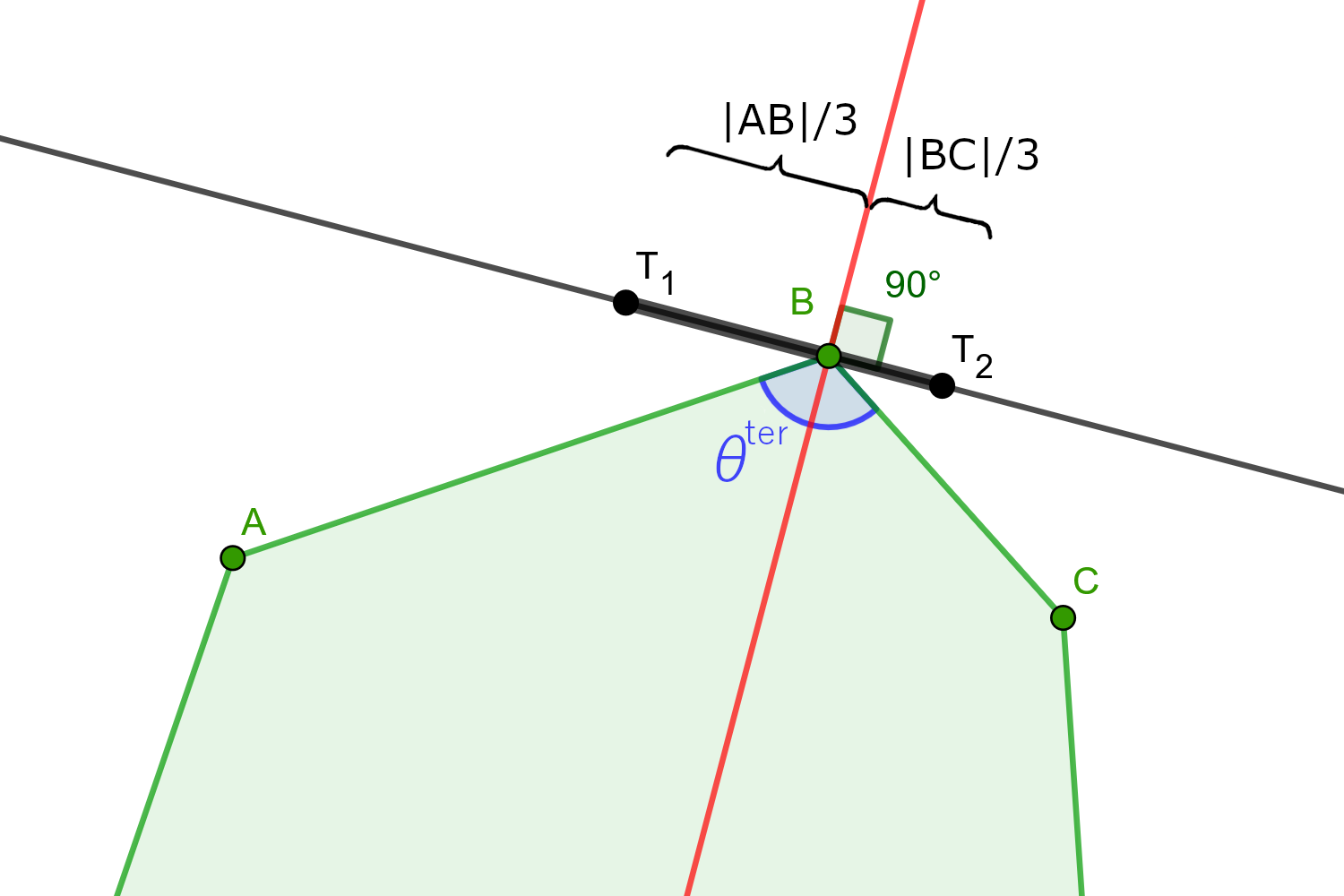}
\caption{Illustration showing how the control points $T_1$ and $T_2$ controlling the tangents of the Bézier curves at point B are determined. The bisector of the tertiary polygon corner with angle $\theta^\text{ter}$ is shown in red.}
\label{fig:tangentpoints}
\end{figure}

Table \ref{table:parameters} summarizes all parameters and their values used in the generation of the microstructures.

\begin{table}[ht]
\centering
\begin{tabular}{|p{0.28\linewidth} | p{0.5\linewidth} | p{0.1\linewidth}|}
\hline
\textbf{Parameter Name} & \textbf{Meaning} & \textbf{Value} \\
\hline
\textbf{Secondary polygon parameters} & & \\
\hline
\verb|MIN_AREA| & Minimum area of secondary polygon & 0.0167 \\
\hline
\verb|MIN_THICKNESS| & Minimum ligament thickness per face & 0.012 \\
\hline
\verb|MAX_REL_THICKNESS| & Maximum ligament thickness per face relative to bisector from straight skeleton & 0.7 \\
\hline
\verb|PROB_RANDOM_FILL| & Probability of bulk material & 0.02 \\
\hline
\textbf{Tertiary polygon parameters} & & \\
\hline
\verb|MIN_RADIUS| & Minimum corner radius & 0.02 \\
\hline
\verb|MIN_D_REL| & Minimum distance of tertiary polygon vertex to secondary polygon vertex, relative to secondary polygon side length & 0.2 \\
\hline
\verb|MIN_SEP_ABS| & Minimum separation between two tertiary polygon vertices on the same secondary polygon edge & 0.02 \\
\hline
\verb|MIN_SEP_REL| & Minimum separation between two tertiary polygon vertices on the same secondary polygon edge, relative to secondary polygon side length & 0.2 \\
\hline
\verb|MIN_AREA2| & Minimum area of tertiary polygon & 0.0083 \\
\hline
\textbf{Mesh parameters} & & \\
\hline
\verb|clmax| & Maximum characteristic element size & 0.041 \\
\hline
\end{tabular}
\caption{Various parameters and their values used in the generation of microstructures.}
\label{table:parameters}
\end{table}

\subsubsection*{Creating the meshes}
The final generated B\'ezier curves serve to define the geometry of the fundamental domain (see Figure \ref{fig:material_generation}i), which was then discretized using Gmsh \cite{Geuzaine2009}. For this dataset, we used quadratic triangular elements with a maximum element size of 0.0408, although Gmsh automatically used smaller elements near fine geometrical features. This size is relative to the area of the unit cell, which is always 1.
The resulting mesh of the fundamental domain was then tiled into the unit cell (see Figure \ref{fig:material_generation}(j)).

We further assumed that a representative volume element (RVE) consisting of 2x2 unit cells would be sufficient to capture the most important pattern transformations (see Figure \ref{fig:material_generation}(k)). In literature, and from experimentation on our part, this does seem to be the case most of the time, although it should be noted that it is possible for a $2\times 2$ RVE to be inadequate in certain load scenarios. For example, Polukhov et al.\cite{Polukhov2018} show that in a electrostatically activated version of the same material a $2\times 3$ RVE is necessary. A global buckling mode may also occur; however, in such case, any RVE size is insufficient, as the deformation wavelength becomes infinite relative to the RVE size and hence full-system (finite-size) analysis would have to be performed. Ideally, Bloch analysis should be used to determine the appropriate RVE size  \cite{Geymonat1993a, Triantafyllidis2006, Boyce2008, Zhang2021a}. Due to its high computational cost, we kept the Bloch analysis outside the scope of our contribution.

\subsection*{Simulating mechanical response}
Each microstructure in the dataset is accompanied by the results of computational homogenization, characterizing the macroscopic response of the entry. For the simulations of the response, we sampled a variety of loading conditions for each geometry, then performed a simulation that captures the large deformation and buckling behavior of the material.
To simulate the response of the material, we performed computational homogenization \cite{Geers2010} using finite element analysis (FEA), taking the applied macroscopic deformation gradient $\bt F_{\mathrm{final}}$ as an input along with the RVE mesh.
The calculation yielded the deformed shape of the RVE and effective macroscopic quantities: the homogenized strain energy density $\mathfrak{W}$, homogenized 1\textsuperscript{st} Piola-Kirchhoff stress tensor $\bt P$ and homogenized 4\textsuperscript{th}-order tangent stiffness tensor $\bt D$.

Given the microstructures' characteristics, the simulations are highly nonlinear as they cover large deformations and especially buckling.
Therefore, we applied the prescribed deformation $\bt F$ in a series of small steps, such that the deformation gradient $\bt F(\tau)$ follows a trajectory parameterized by the pseudo-time $\tau$, which attains values from 0 to 1, with $\bt F(0) = \bt I$ and $\bt F(1) = \bt F_{\mathrm{final}}$. We adaptively changed the step length to facilitate robust simulations.
Consequently, the effective quantities are then also functions of $\tau$, i.e., $\mathfrak{W}(\tau)$, $\bt P(\tau)$ and $\bt D(\tau)$. Through this type of displacement control, we can handle snap-through, but not snap-back \cite[\S 5.1.5, p.165]{wriggers2008nonlinear}.

\label{par:bulk_material}
For the bulk material, we used the hyperelastic Bertoldi-Boyce constitutive model under the plane strain assumption\cite{Boyce2008}, which is defined by the following microscopic strain energy density $\mathfrak{W}_\text{m}(\bt F_\text{m})$:
\begin{equation}
    \mathfrak{W}_\text{m}(\bt F_\text{m}) = {c_1} \left(I_1 - 2\right) + {c_2} \left(I_1 - 2\right)^2 - 2{c_1} \ln\left(J_\text{m}\right) + \frac{{K}}{2} \left(J_\text{m}-1\right)^2.\label{eq:bertoldiboyce}
\end{equation}
Here, $J_\text{m} = \det\bt F_\text{m}$ and $I_1 = \tr \bt C_\text{m}$ is the first invariant of the right Cauchy-Green tensor $\bt C_\text{m} = \bt F_\text{m}^T \cdot \bt F_\text{m}$. Values of the material parameters were chosen in accordance with Boyce \& Bertoldi\cite{Boyce2008} as ${c_1}=\SI{0.55}{MPa}$, ${c_2}=\SI{0.3}{MPa}$, ${K}=\SI{55}{MPa}$.
In the 3D case, $I_1-3$ in place of $I_1-2$ has to be used, as in the original Bertoldi-Boyce paper.

\subsubsection*{First-order Computational homogenization}
\label{par:governing_equation}
The balance equation, neglecting any inertia effects, reads as
\begin{equation}
    \vec{\nabla}_{{0},\text{m}} \cdot \bt P_\text{m}^{{T}}(\bt F_\text{m}(\vec{X})) = \vec 0 \quad \forall \vec{X} \in \Omega_0 \label{eq:PDE},
\end{equation}
which was solved for the deformed microscopic positions $\vec{x}$. Here,  $\Omega_0$ denotes the RVE domain, and $\vec{\nabla}_{0,\text{m}}\cdot$ denotes the divergence of the second order tensor field $\bt P_\text{m}$ with respect to the initial microstructural position $\vec{x}_0$. We use the following definition of the $\vec{\nabla}_0$ operator:
\begin{equation}
    \vec{a} = \vec{\nabla}_0 \cdot \bt A
    \quad \Longleftrightarrow \quad
    a_{j} = \sum_i \pdv{A_{ij}}{x_{0,i}}.
\end{equation}

\label{par:homogenized_quantities}
We adopted the classical kinematical ansatz of first-order computational homogenization, expressed as:
\begin{equation}
    \vec{x}(\vec{x}_0) = \bt F \cdot \vec{x}_0 + \vec w(\vec{x}_0), \quad \forall \vec{x}_0 \in \Omega_0,
    \label{eq:kinematics}
\end{equation}
where $\vec w (\vec{x}_0)$ is a periodic fluctuation field, which sufficed to ensure
\begin{equation}
    \bt F = \frac{1}{|\Omega_0|} \int_{\Omega_{0}} (\vec\nabla_0 \vec{x})^T \dd{\vec{x}_0}.
\end{equation}
The fluctuation field $\vec w(\vec{x}_0)$ was obtained by minimizing the strain energy of the RVE, which corresponds to ensuring the microscale equilibrium equation given by Equation \eqref{eq:PDE} was satisfied in a weak sense.

In the numerical implementation, $\bt F$ acted as a directly prescribed load. The periodicity of $\vec w (\vec{x}_0)$ was enforced using a periodic projection of unique degrees of freedom, i.e., the image-source approach.
In addition, we fixed one boundary node in place to eliminate rigid-body modes.

We solved for $\vec w$ incrementally, using the standard Newton-Raphson method combined with a line search. After each load increment, we checked the stability of a converged state by evaluating the eigenvalues of the tangent stiffness matrix. If instability was identified, we went back to the previous time step and tried again with a smaller load increment, i.e., a smaller $\Delta \tau$. If $\Delta \tau$ fell below a predefined threshold (0.005, sometimes lower for the more tricky simulations), we assumed that the system was close to the critical point. We then perturbed the current state in a direction of the eigenmode corresponding to the smallest eigenvalue and let the system relax.

Once a stable solution for $\vec w$ was obtained, the homogenized $\mathfrak{W}$ and $\bt P$ were obtained as volume averages of their microscopic counterparts, i.e.,
\begin{align}
    \mathfrak{W} &= \frac{1}{\left|\Omega_0\right|}\int_{\Omega_{0}} \mathfrak{W}_\text{m} \left(\bt F_\text{m} \left(\vec{x}\right)\right)\dd \vec{x}_0 \label{eq:homoW},\\
    \bt P &= \pdv{\mathfrak{W}}{\bt F} = \frac{1}{\left|\Omega_0\right|}\int_{\Omega_{0}} \bt P_\text{m} \left(\bt F_\text{m} \left(\vec{x}\right)\right)\dd \vec{x}_0\label{eq:homoP},
\end{align}
where $\left|\Omega_0\right|$ denotes the volume of the RVE in the reference configuration.
The consistent homogenized stiffness $\bt D$ is defined as \cite[p. 227]{Tadmor2012}:
\begin{equation}
    {}^4 \bt D = \frac{\partial^2{\mathfrak{W}}}{\partial{\bt F}\partial{\bt F}} = \pdv{\bt P}{\bt F}, \label{eq:homoD}
\end{equation}
or equivalently
\begin{equation}
    \delta \bt P = {}^4 \bt D : \delta \bt F,
\end{equation}
where the double-dot product between a 4\textsuperscript{th} and 2\textsuperscript{nd} order tensor is defined as
\begin{equation}
    \bt C = \bt {}^4 \bt A : \bt B \quad \Leftrightarrow \quad C_{ij} = \sum_{k,l} A_{ijkl} B_{lk}.
\end{equation}
Tensor ${}^4 \bt D$ was obtained via static condensation via static condensation of the fluctuation degrees of freedom from an extended tangent stiffness matrix, which incorporates both the fluctuation field and the generalized degrees of freedom associated with the prescribed macroscopic deformation gradient $\bt F$. This procedure follows the approaches of, e.g., Kouznetsova, 2002\cite[\S 2.4.4]{Kouznetsova2002} and Miehe, 2003 \cite{Miehe2003}, simplified in our case by the explicit separation of degrees of freedom pertinent to the kinematic ansatz in Equation~\eqref{eq:kinematics}.

The FE simulations were performed using an in-house MATLAB code which is available at \url{https://github.com/FHendriks11/mechmetamat_homogenization}.

\subsubsection*{Sampling of the Deformation Gradient}
Per geometry we sampled the RVE's response to 12 different prescribed loading paths, each defined by a different prescribed final macroscopic deformation gradient $\bt F$.
Using the right polar decomposition, $\bt F$ can be decomposed into the symmetric macroscopic right-stretch tensor $\bt U$ and a rotation tensor $\bt R$, i.e., $\bt F = \bt R \cdot \bt U$.
Because $\bt R$ encodes rotation, which does not meaningfully affect the results, we only sampled symmetric $\bt F$, which corresponds to prescribing the right-stretch tensor $\bt U$.

The sampling was carried out using the strategy introduced by Kunc and Fritzen\cite{Kunc2019}, which samples a large variety of symmetric positive definite tensors in 2D. Each sample $bt U$ is defined by three independent parameters: volumetric deformation given by $J=\det \bt U$, the orientation of the principal strains parameterized by the angle $\phi$, and the deviatoric magnitude $t$ of the deformation.

The right stretch tensor $\bt U$ (and from it $\bt F$) in its matrix-representation $\underline{\underline{U}}$ was then constructed as follows:
\begin{equation}
    \underline{\underline{U}} = J^{1/2}
    \exp\left(
        t \left(
            \alpha \underline{\underline{Y}}^{(1)} + \beta \underline{\underline{Y}}^{(2)}
        \right)
    \right),
\end{equation}
with
\begin{equation}
    \underline{\underline{Y}}^{(1)} = \sqrt{\frac{1}{2}}
    \begin{bmatrix}
        1 & 0 \\
        0 & -1 \\
    \end{bmatrix} \quad \text{and} \quad
    \underline{\underline{Y}}^{(2)} = \sqrt{\frac{1}{2}}
    \begin{bmatrix}
        0 & 1 \\
        1 & 0 \\
    \end{bmatrix},
\end{equation}
where $\alpha = \sin \phi$ and $\beta = \cos \phi$.
For our sampling, we set $J \in [J_{\textrm{min}}, J_{\textrm{max}}]$, $t \in [0, t_{\textrm{max}}]$ and $\phi \in [0, 2\pi]$. This sampling results in an isotropic distribution of principal directions.

Since all simulations were performed in load increments anyway, we sampled only the extreme values; i.e., $J=J_{min}$, $J=J_{max}$ or $t=t_{max}$. Because we performed 12 simulations for each geometry, we sampled four values of $\bt F$ with $J=J_{\textrm{min}}$, four with $J=J_{\textrm{max}}$ and four with $t=t_{\textrm{max}}$.
We chose $J_{\textrm{min}}$ equal to the volume fraction of the geometry whose response we were computing, because at that point, virtually all of the deformed configurations include self-contact. Continuing beyond that point is not physically meaningful, since our simulations did not include contact.
For $J_{\textrm{max}}$, we chose 1.5, and $t_{\textrm{max}}$ was set to 0.5.
We then sampled four combinations of the other two parameters (either $t$ and $\phi$ if $J$ was at an extremum, or $J$ and $\phi$ if $t$ was at its maximum), by dividing the range of these parameters into two equal intervals and sampling with uniform probability from each combination of these intervals.

\begin{figure}[htbp]
\centering
\includegraphics[width=\linewidth]{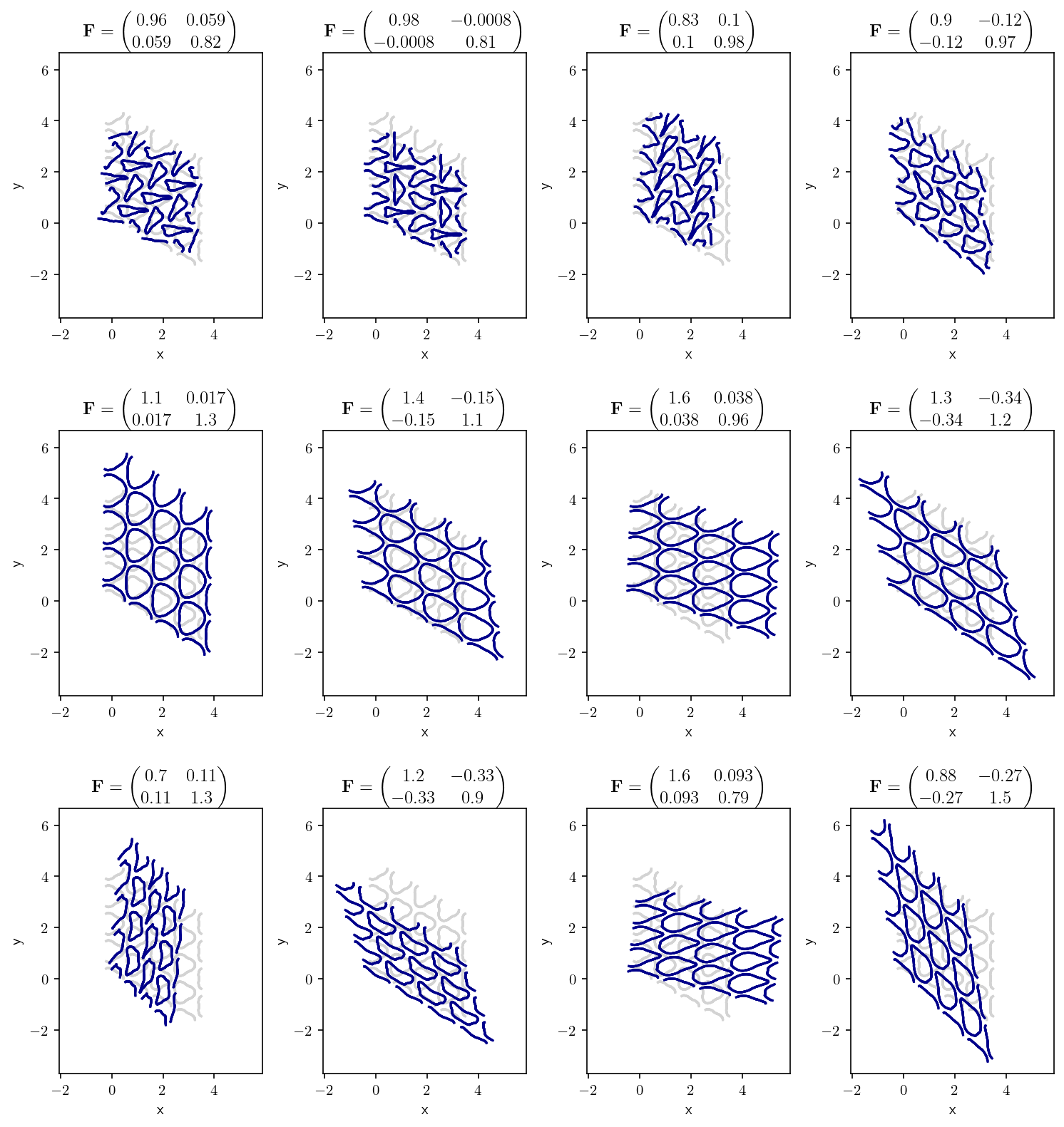}
\caption{Deformed configurations (in blue) at the end of the 12 simulations of the same geometry with \emph{p3m1} symmetry. The undeformed geometry is shown in grey. To allow for better understanding of the pattern transformation, the entire RVE is plotted 4 times per figure (2 repetitions along each lattice vector). The title of each subfigure indicates the deformation gradient $\bt F$ at the end of the simulation.}
\label{fig:deformations}
\end{figure}

\begin{figure}[htbp]
\centering
\includegraphics[width=\linewidth]{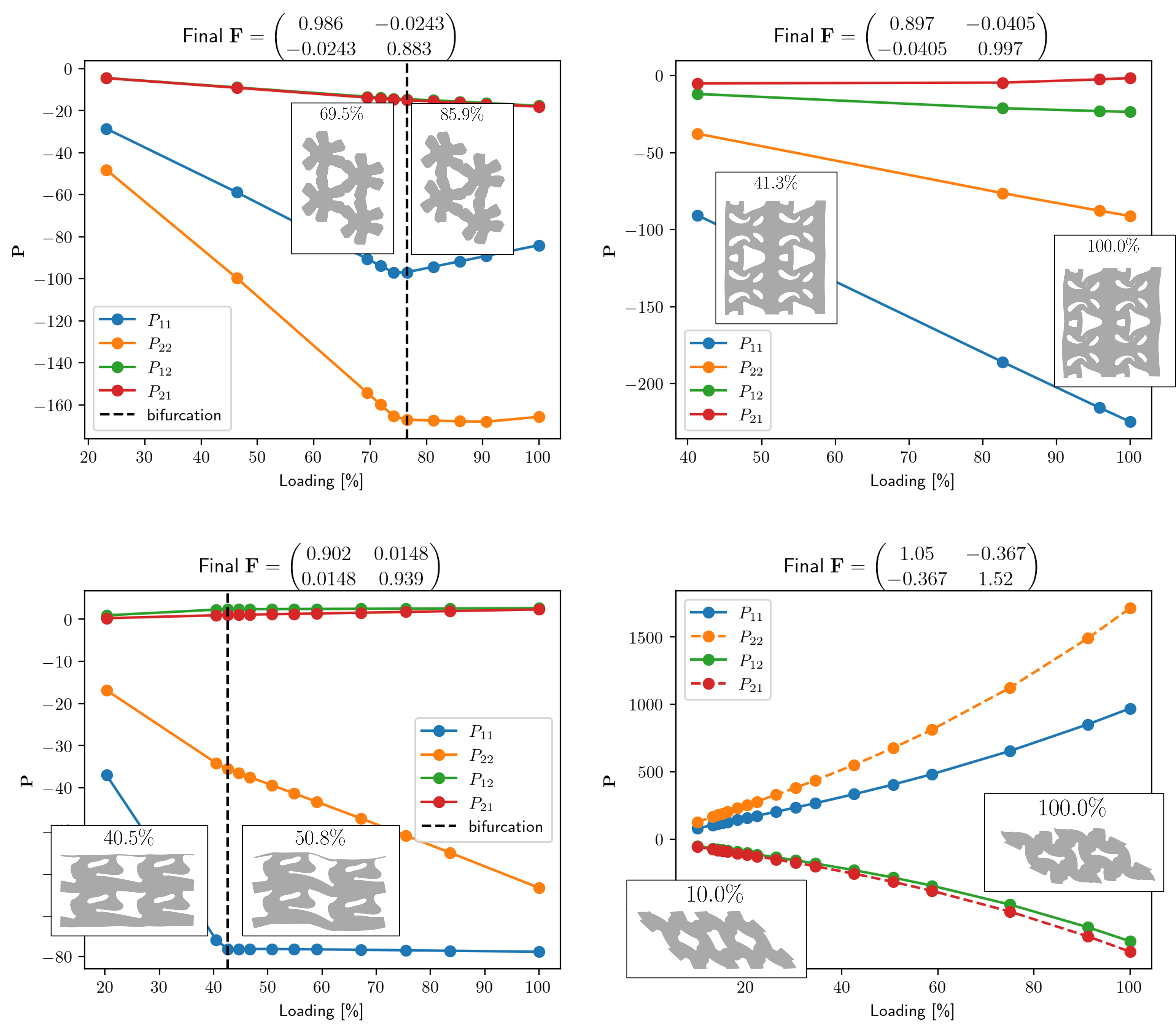}
\caption{\textcolor{mygreen}{1\textsuperscript{st} Piola-Kirchhoff stress tensor components vs percentage loading over the course of a trajectory. Each figure represents the stress response of one material under the loading condition applied to it.}}
\label{fig:stress_loading}
\end{figure}

An example of the resulting set of deformations at the end of all twelve simulations for one geometry is provided in Figure \ref{fig:deformations}\textcolor{mygreen}{, and examples of the resulting stress over the course of four different trajectories (each of a different geometry) are given in Figure \ref{fig:stress_loading}}.

\section*{Data Records}
The data set, available in a Zenodo repository (\url{https://zenodo.org/records/15849549}), contains 1,020 microstructural geometries in total: 60 distinct ones for each of the 17 wallpaper groups \cite{datasetzenodo}.
The dataset therefore consists of 1,020 .pkl and 1,020 .png files, which are placed into the subdirectories \verb|pickle_files\| and \verb|images\|. Each .pkl file contains data describing the geometry of one microstructure and the results of the 12 simulations of this geometry. Each .png file then contains a visualization of the geometry. The total size of the .pkl files is 47.2 GB, and 87.6 MB of .png files, making the total size of the dataset 47.3 GB.

Since we performed 12 simulations for each geometry, our dataset contains $12 \times 60 \times 17 = 12,240$ trajectories, each of which has on average 11.1 pseudo-time steps. Considering each pseudo-time step to be a separate load case, the data set contains 135,947 microstructural responses to a prescribed $\bt F$ in total.

\subsection*{.pkl file}
\begin{figure}[htbp]
\centering
\includegraphics[width=\linewidth]{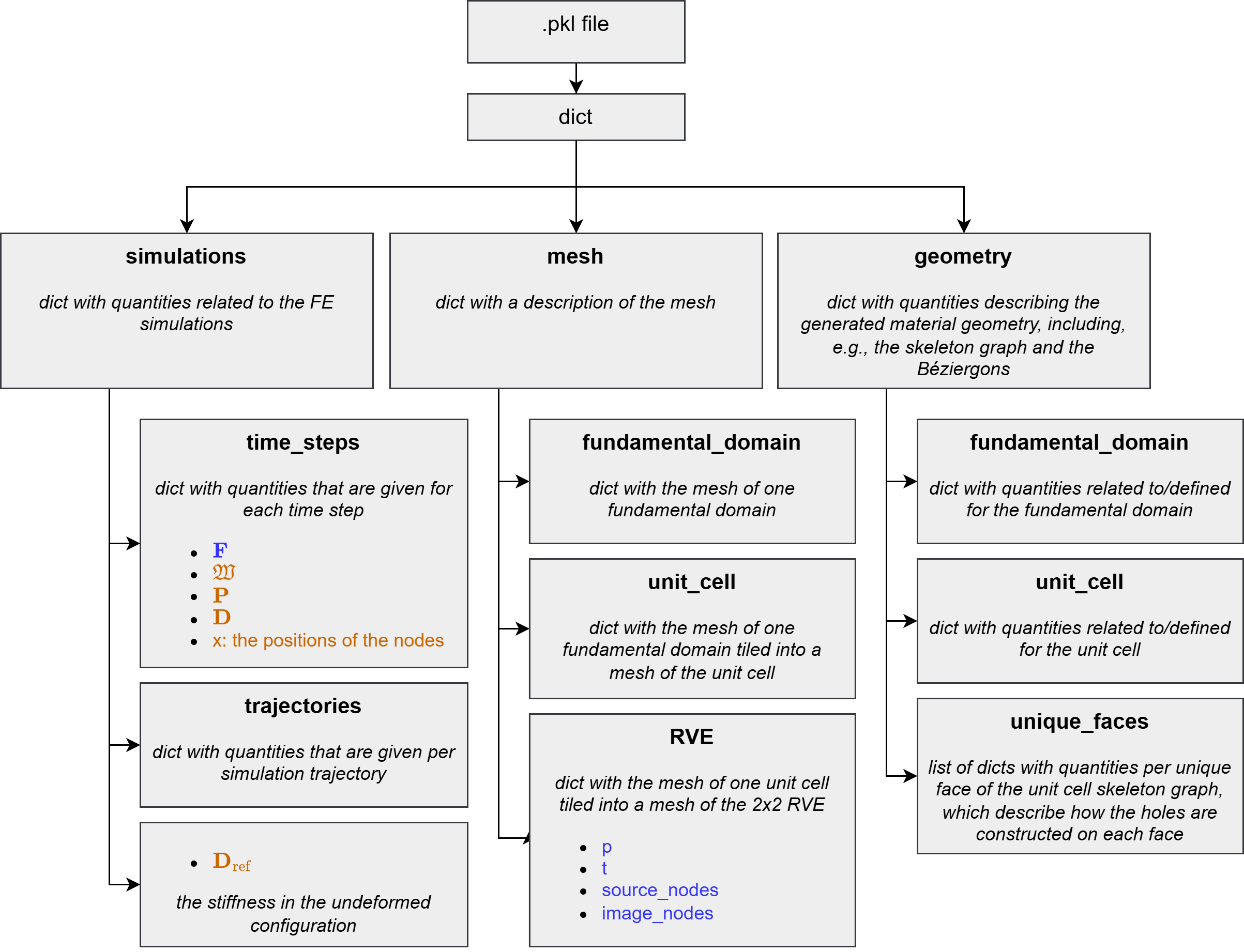}
\caption{Hierarchical representation of the structure of the .pkl files. Quantities in blue are the inputs to the simulations, and quantities in orange are the outputs.}
\label{fig:pkl_file_structure}
\end{figure}

Each .pkl file contains a Python dictionary (dict), which contains multiple keys, whose values are dicts themselves. Figure \ref{fig:pkl_file_structure} shows the structure of the dicts inside the .pkl file. The README markdown file in the Zenodo contains an exhaustive list of all the quantities in this dict and their meanings. The most important quantities are the inputs and outputs of the simulations, which are listed here.\\
Inputs:
\begin{itemize}
    \item \verb|data['simulations']['time_steps']['F']|: a numpy array of float64 values of shape (number of time steps, 2, 2), containing the deformation gradient $\bt F$ at each time step.
    \item \verb|data['mesh']['RVE']['p']|: a numpy array of float64 values of shape (number of nodes, 2), containing the $x,y$-coordinates of the nodes of the RVE mesh.
    \item \verb|data['mesh']['RVE']['t']|: a numpy array of int32 values of shape (number of elements, 6), containing the indices of the nodes of each triangular quadratic element.
    \item \verb|data['mesh']['RVE']['source_nodes']|: list of uint16 values of indices of the boundary nodes that the \verb|periodic_image_nodes| are dependent on.
    \item \verb|data['mesh']['RVE']['image_nodes']|: list of uint16 values of indices of boundary nodes whose displacements are not solved for but instead calculated from their corresponding source node, in order to ensure periodic boundary conditions.
\end{itemize}
Outputs:
\begin{itemize}
    \item \verb|data['simulations']['time_steps']['W']|: a numpy array of float64 values of shape (number of time steps, ), containing the strain energy density $\mathfrak{W}$ at each time step.
    \item \verb|data['simulations']['time_steps']['P']|: a numpy array of float64 values of shape (number of time steps, 2, 2), containing the first Piola-Kirchhoff stress tensor $\bt P$ at each time step.
    \item \verb|data['simulations']['time_steps']['D']|: a numpy array of float64 values of shape (number of time steps, 2, 2, 2, 2), containing the tangent stiffness $\bt D$ at each time step.
    \item \verb|data['simulations']['D_ref']|: a numpy array of float64 values of shape (2, 2, 2, 2), containing the tangent stiffness $\bt D$ in the reference (undeformed) configuration.
    \item \verb|data['simulations']['time_steps']['x']|: a numpy array of float64 values of shape (number of time steps, number of nodes, 2), containing the $x,y$-coordinates of the nodes of the RVE at each time step.
\end{itemize}

\subsection*{.png file}
Each .png file contains a visualization of the geometry. The image always captures the whole RVE, of which one unit cell is indicated by thick black lines and a light grey background. In this unit cell, the mesh is shown as triangles of various colors. Within the unit cell, one fundamental domain is indicated, also with thick black lines and with a slightly darker grey background. Examples from all 17 wallpaper groups are shown in Figures \ref{fig:examples_p1_p2}, \ref{fig:examples_pm_pg_cm}, \ref{fig:examples_pmm_pmg_pgg_cmm} and \ref{fig:examples_p4_p4m_p4g_p3_p3m1_p31m_p6m_p6}.

\section*{Technical Validation}
We thoroughly validated the microstructure generation algorithm, the discretization (checking both the validity of the mesh as well as the mesh refinement) and the simulations.
The microstructure generation algorithm produces connected and periodic geometries. Both features are rigorously tested in the subsequent steps: 
periodicity is checked as a side effect when stitching together multiple copies of the fundamental domain to create the unit cell, and later 4 copies of the unit cell to create the RVE; if periodicity were not respected, this stitching together triggers an error. Similarly, if the geometry were not connected, the numerical simulations would result in an error.

Sometimes the generation algorithm fills too many graph faces, resulting in a very high volume fraction, which hampers buckling. For this reason, we discard geometries with a volume fraction above 0.75.

We prescribe the minimum thickness of the ligaments by controlling the distance between corners of the initial and secondary face polygon. However, since the final B\'ezier curves that specify the ligaments can `bulge' out beyond the boundaries of the secondary face polygon, we added an extra check in the code that computes the distance between B\'ezier curves and discards the geometry if this distance is too small, ensuring the minimum thickness of the ligaments.

Very rarely, Gmsh generates invalid meshes with slightly overlapping elements. Therefore, the fundamental domain mesh was validated by checking for elements overlapping other elements or themselves. This was done by checking for crossing element edges. If any were found, the geometry was discarded.

\label{subsec:mesh_convergence}
To assess the necessary level of mesh refinement, we perform a mesh convergence study on 4 different geometries, containing a variety of features, including very thin ligaments and multiple holes. We then assess the accuracy of the calculations and their convergence.

\begin{figure}[htbp]
    \centering
    \begin{minipage}[t]{0.49\textwidth}
        \centering
        \includegraphics[width=\textwidth]{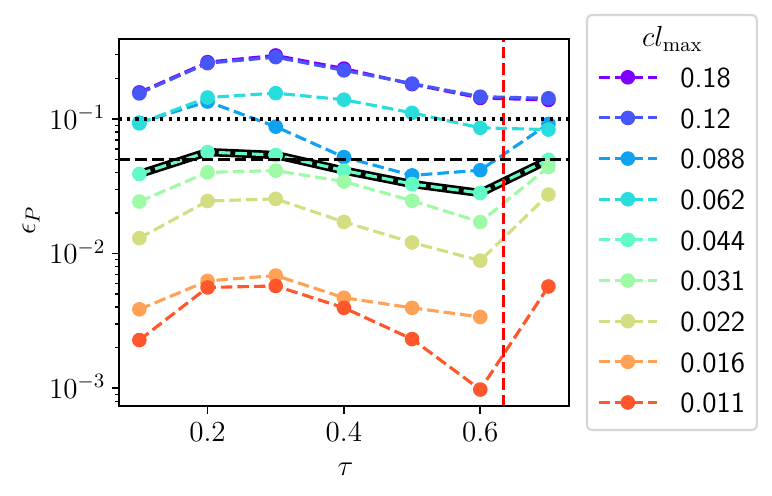}
        \caption*{(a)}  
    \end{minipage}
    \hfill
    \begin{minipage}[t]{0.49\textwidth}
        \centering
        \includegraphics[width=\textwidth]{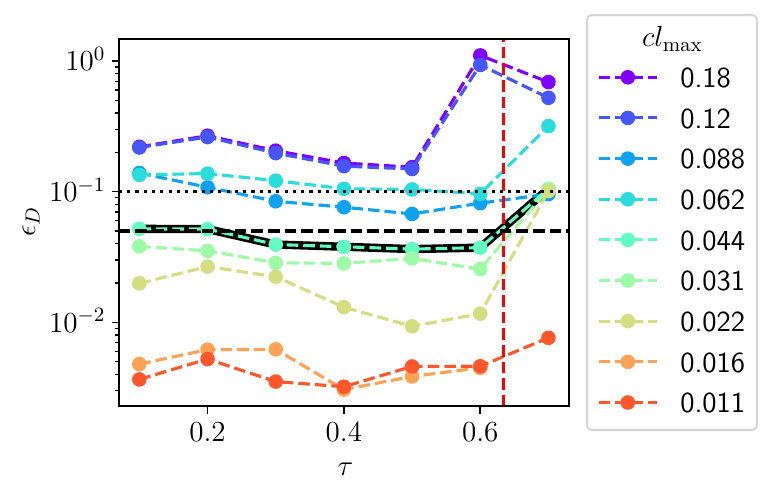}
        \caption*{(b)}  
    \end{minipage}
    \vspace{0.5cm}
    \begin{minipage}[t]{0.5\textwidth}
        \centering
        \includegraphics[width=\textwidth]{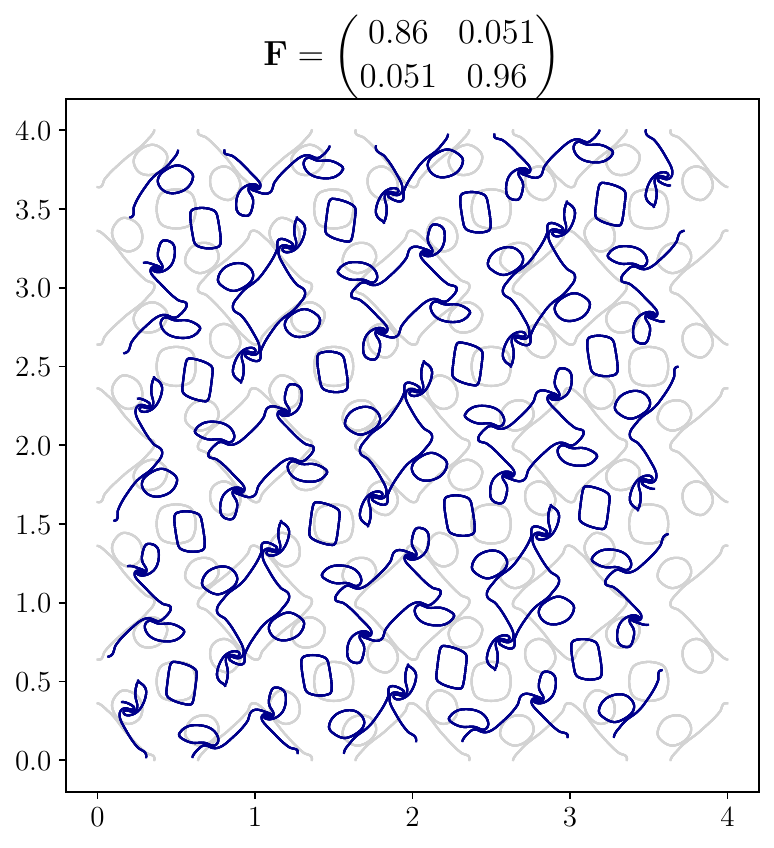}
        \caption*{(c)}  
    \end{minipage}
    \caption{Relative error in (a) the homogenized 1\textsuperscript{st} Piola-Kirchhoff stress tensor $\bt P$ and (b) tangent stiffness $\bt D$ for a geometry with \emph{p4} symmetry, over the course of the simulation until contact, i.e., for $\tau \in [0, 0.70]$, for varying maximum characteristic element sizes $cl_{\mathrm{max}}$. The $cl_{\mathrm{max}}=0.044$ closest to our final chosen maximum characteristic element size of $0.0408$ is additionally indicated with a black line. Subfigure (c) shows the geometry of the test RVE, with the undeformed geometry in grey and the deformed geometry at the end of the simulation in blue. To allow for better understanding of the pattern transformation, the entire RVE is plotted 4 times (2 repetitions along each lattice vector). The title of subfigure (c) indicates the deformation gradient $\bt F$ at the end of the simulation when contact happens (at $\tau=0.70$).}
    \label{fig:mesh_convergence_p4}
\end{figure}

For each geometry, we use a different applied deformation gradient $\bt F$ with $J=0.75$, i.e., with significant volumetric compression, which are the most difficult type of simulations prone to involve buckling. We then calculate the relative error $\epsilon_P$ in the 1\textsuperscript{st} Piola-Kirchhoff stress tensor $\bt P$ and $\epsilon_D$ in the tangent stiffness $\bt D$, compared to the mesh with the smallest element size. $\epsilon_P$ and $\epsilon_D$ are calculated as
\begin{equation}
    \epsilon_P = \frac{\lVert \bt P^\tau - \bt P^\tau_{\textrm{fine}} \rVert}{\lVert \bt P^\tau_{\textrm{fine}} \rVert}
\end{equation}
and
\begin{equation}
    \epsilon_D = \frac{\lVert \bt D^\tau - \bt D^\tau_{\textrm{fine}} \rVert}{\lVert \bt D^0_{\textrm{fine}} \rVert},
\end{equation}
where $\lVert \cdot \rVert$ is the Frobenius norm, and $\tau$ is the pseudo-time. The subscript `fine' indicates it is the finest mesh, for which the maximum characteristic element size $cl_{\mathrm{max}}=0.0078$. Note that for $\epsilon_P$, we normalize against $\lVert \bt D^0_{\textrm{fine}} \rVert$ instead of $\lVert \bt D^\tau_{\textrm{fine}} \rVert$, because the latter can attain very small values or even zero at the buckling point. The convergence study demonstrates that, for a mesh size of $\mathtt{cl\_max}=0.044$, both $\epsilon_P$ and $\epsilon_D$ are below 5\% most of the time, even for these most challenging cases, and consistently below 10\%, with the sole exception of $\epsilon_D$ close to the buckling point. At that point, there is a discontinuity in $\bt D$, and its location can shift with varying mesh sizes. Since a 5\% to 10\% accuracy is good enough for our purposes (neural network training and properly identifying the qualitative response of the material), we deem $\mathtt{cl\_max} = 0.044$ or lower acceptable.
Hence, the adopted discretisation represents an acceptable compromise between the accuracy of the reported responses and the computational cost of generating the dataset.
See Figure \ref{fig:mesh_convergence_p4} for the relative error in $\bt P$ and $\bt D$ for one of the test geometries, over the course of the simulation.

Technical validation of the simulation code itself was carried out by comparing its results to the literature for two common geometries; the square stacked holes and the hexagonally stacked holes. Our code correctly reproduces the same pattern transformations as seen in the literature \cite{Rokos2020}.

Since our simulation code does not take into account contact, we check for contact in the results during postprocessing, by checking for overlapping element edges. Cases with overlapping elements are removed, such that only the part of the trajectory before contact remains in the dataset.

\section*{Usage Notes}
A Jupyter notebook titled \verb|Usage_notes.ipynb|, which is provided in the Zenodo repository, demonstrates use cases of the dataset. It shows how to load the data, how to access the different quantities in the data, and how to visualize the geometries. For example, it visualizes the mesh before and after deformation, and plots the homogenized quantities $\bt P$ and $\bt D$ as a function of the pseudo-time $\tau$.

The envisioned use cases for the dataset are at least twofold: training machine learning models and studying symmetries in mechanical metamaterials.

For the purpose of training machine learning models, we suggest using the data to create a graph representation suitable for a graph neural network, by removing at least the bulk nodes from the mesh and creating a graph from the remaining boundary nodes, as suggested by Hendriks et al\cite{Hendriks2025}, or from the Bézier control points. Additionally, the graph skeleton used to generate the geometry can be used in the graph representation.

For the purpose of studying how symmetries in a material relate to its mechanical properties, one can investigate the homogenized quantities for each symmetry in search for patterns in the dataset. \textcolor{mygreen}{For example, one can examine the auxeticity per wallpaper group.
Auxeticity refers to a negative Poisson's ratio, where a material expands perpendicular to the direction of stretching and contracts perpendicular to the direction of compression.
The Poisson's ratio can be calculated from the tangent stiffness tensor $\bt D$. In our dataset, we find that in the reference configuration, 306 out of the 1,020 geometries are auxetic in at least one direction, of which 110 geometries are auxetic in all directions. Code to plot the Poisson's ratio as a function of direction is provided in the} \verb|Usage_notes.ipynb| \textcolor{mygreen}{notebook, where the anisotropic Poisson's ratio is calculated as \cite{gorodtsov2019extreme}:
\begin{align}
    \nu(\vec{n}, \vec m) &=
    -\frac{1}{E(\vec n)}s_{ij\alpha\beta} n_i n_j m_\alpha m_\beta,\\
    E(\vec n) &= s_{ij\alpha\beta} n_i n_j n_\alpha n_\beta,
\end{align}
with $\vec n$ a unit vector in the loading direction and $\vec m$ a unit vector perpendicular to $\vec n$ (only two directions possible in 2D, which will give the same result), and $s_{ij\alpha\beta}$ the components of the compliance tensor $\bt S$, which is the pseudo-inverse of the stiffness tensor $\bt D$.}

\begin{figure}[htbp]
\centering
\includegraphics[width=0.6\linewidth]{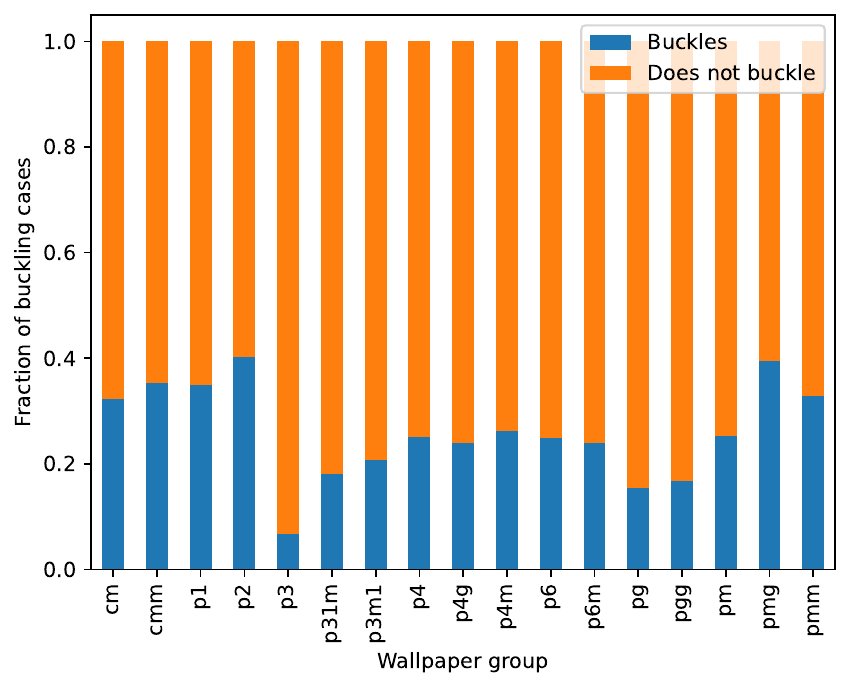}
\caption{Fraction of trajectories that buckled for each wallpaper group, calculated as the number of trajectories that buckled divided by the total number of trajectories for that wallpaper group.}
\label{fig:bucklingfraction}
\end{figure}

One can also look at the tendency of each wallpaper group to generate materials that buckle, and which buckling modes show up. For example, Figure \ref{fig:bucklingfraction} shows the fraction of geometries that buckled for each wallpaper group, indicating large differences between the wallpaper groups in their tendency to buckle.

\section*{Code availability}

The code used to generate the geometries is available at \url{https://github.com/FHendriks11/wallpaper_microstructures}. The code used to simulate the mechanical response is available at \url{https://github.com/FHendriks11/mechmetamat_homogenization}. Both repositories are avilable under the CC-BY 4.0 license.

We used Python 3.9.18, with the following package version\textcolor{mygreen}{s}: scipy 1.13.1, scikit-learn 1.3.0 (solely for the \verb|radius_neighbors_graph| for fast deduplication of points and checking of mesh or B\'ezier curve overlap), scikit-geometry 0.1.2 (solely for the straight skeleton), NumPy 1.26.4, meshio 5.3.4 (solely for reading gmsh mesh files) and Matplotlib 3.9.2. We used Gmsh 4.12.2 to generate the meshes, and MATLAB R2021a to perform the simulations.
\textcolor{mygreen}{
\noindent These packages and software can be found at the following URLs:
\begin{itemize}
    \item Python: \url{https://www.python.org/}
    \item SciPy: \url{https://scipy.org/}
    \item scikit-learn, radius\_neighbors\_graph: \url{https://scikit-learn.org/1.3/modules/generated/sklearn.neighbors.radius_neighbors_graph.html}
    \item scikit-geometry, straight skeleton: \url{https://scikit-geometry.github.io/scikit-geometry/skeleton.html}
    \item NumPy: \url{https://numpy.org/}
    \item meshio: \url{https://github.com/nschloe/meshio}
    \item Matplotlib: \url{https://matplotlib.org/}
    \item Gmsh: \url{https://gmsh.info/}
    \item MATLAB: \url{https://www.mathworks.com/products/matlab.html}
\end{itemize}
}

\section*{Author contributions statement}
F.H. conceived the method of generating new microstructures and implemented and validated this, adapted the MATLAB simulation code and performed the MATLAB simulations and wrote the initial version of the manuscript. M.D. wrote the MATLAB simulation code. V.M., M.D., M.G.D.G., and O.R. advised, acquired funding, and supervised. K.V. helped with the periodic graph connectivity algorithm and consulted on aspects regarding computational geometry. All authors reviewed the manuscript.

\section*{Competing interests} 

The authors declare no competing interests.

\section*{Acknowledgements} 
This project has received funding from the Eindhoven Artificial Intelligence Institute (EAISI).
MD's work was supported by the Czech Science Foundation through Project No. 19-26143X in 2023 and by the European Union under the ROBOPROX project (reg. no. CZ.02.01.01/00/22 008/0004590) since 2024.

\appendix
\section{Periodic graph connectivity check}
\label{sec:appconn_check}
\begin{figure}[htbp]
\centering
\includegraphics[width=0.4\linewidth]{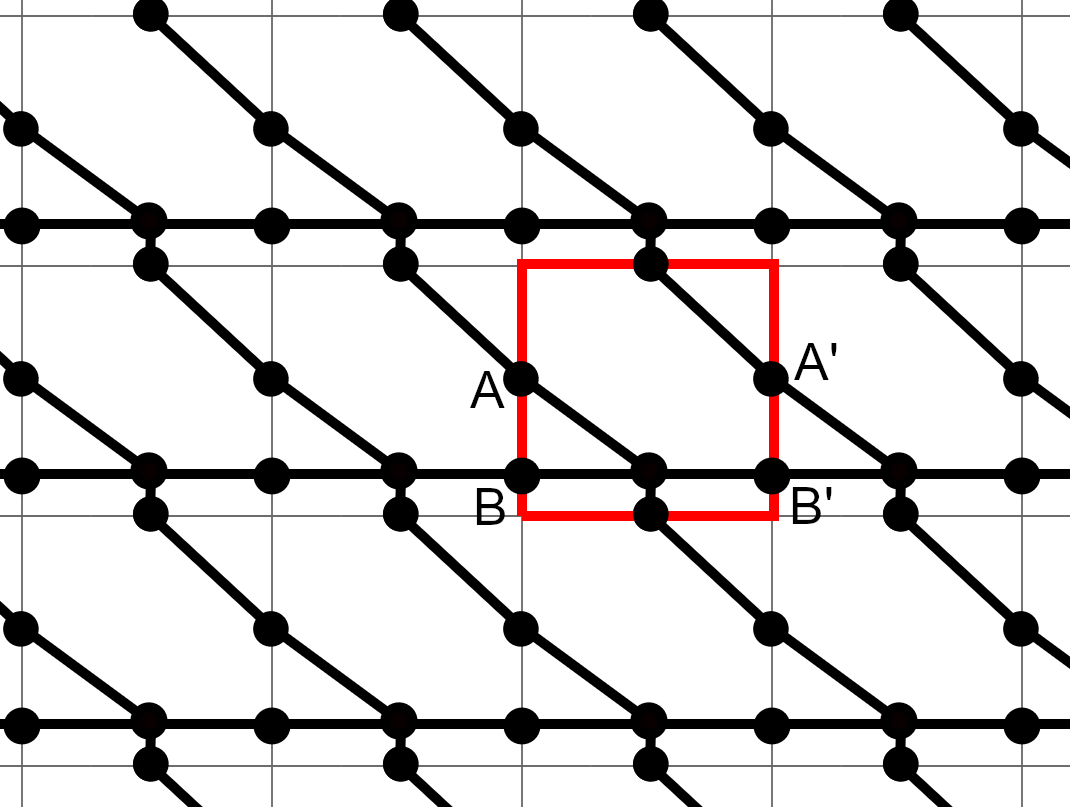}
\caption{A periodic graph of wallpaper group \emph{p1}, with the fundamental domain (equal to the unit cell in this case) outlined in red. Points $A$ and $B$ are connected by the graph inside the fundamental domain, but their periodic equivalent points $A'$ and $B'$ are not.}
\label{fig:connectivityalgorithm}
\end{figure}

Checking the connectivity of a periodic graph is not as simple as checking connectivity of the fundamental domain graph; a connected fundamental domain graph is neither necessary nor sufficient for a connected infinitely tiled graph. For example, Figure \ref{fig:connectivityalgorithm} shows a periodic graph which is not connected inside the fundamental domain, but its infinitely tiled graph \emph{is} connected. To assess the connectivity of the infinite graph we do the following:

\begin{enumerate}
    \item Make sure the fundamental domain shape has the smallest number of vertices possible. For example, the symmetry group \emph{pmg} allows for either a triangular or a rectangular fundamental domain, so we opt for the triangle. If a rectangular fundamental domain was used when a triangular one was possible, convert it to a triangular one.
    \item Check if every boundary has at least one point on it which has at least one edge connected to it. If not, the graph is not connected.
    \item Check if the fundamental domain graph is connected.
    \item If yes, the graph is connected.
    \item If yes to (1), but no to (2), we check if there are equivalent pairs of nodes that are connected on one boundary but not on another. For example, if there are two nodes $A$ and $B$ on the same fundamental domain boundary, and this boundary is equivalent to another boundary because of the symmetry of the tiling, then on that equivalent boundary there must be an equivalent pair of nodes $A'$ and $B'$ (see Figure \ref{fig:connectivityalgorithm}). Then if $A$ is connected to $B$, but $A'$ is not connected to $B'$ on this fundamental domain, then in the infinitely tiled graph, $A'$ and $B'$ will also be connected, but through another copy of the fundamental domain.
    \item If we find such a pair of nodes $A'$ and $B'$, add an edge between them, and check again if the fundamental domain is connected.
    \item Repeat from (2) until there are no such pairs.
    \item If the fundamental domain graph is still not connected, the periodic graph is not connected.
\end{enumerate}

\begin{thebibliography}{10}
\urlstyle{rm}
\expandafter\ifx\csname url\endcsname\relax
  \def\url#1{\texttt{#1}}\fi
\expandafter\ifx\csname urlprefix\endcsname\relax\def\urlprefix{URL }\fi
\expandafter\ifx\csname doiprefix\endcsname\relax\def\doiprefix{DOI: }\fi
\providecommand{\bibinfo}[2]{#2}
\providecommand{\eprint}[2][]{\url{#2}}

\bibitem{GuellIzard2020}
\bibinfo{author}{{Guell Izard}, A.} \& \bibinfo{author}{Valdevit, L.}
\newblock \bibinfo{journal}{\bibinfo{title}{{Magnetoelastic Metamaterials for
  Energy Dissipation and Wave Filtering}}}.
\newblock {\emph{Advanced Engineering Materials}}
  \textbf{\bibinfo{volume}{22}}, \bibinfo{pages}{1901019},
  \url{https://doi.org/10.1002/adem.201901019} (\bibinfo{year}{2020}).

\bibitem{Ning2021}
\bibinfo{author}{Ning, S.} \emph{et~al.}
\newblock \bibinfo{journal}{\bibinfo{title}{{The role of material and geometric
  nonlinearities and damping effects in designing mechanically tunable acoustic
  metamaterials}}}.
\newblock {\emph{International Journal of Mechanical Sciences}}
  \textbf{\bibinfo{volume}{197}}, \bibinfo{pages}{106299},
  \url{https://doi.org/10.1016/j.ijmecsci.2021.106299} (\bibinfo{year}{2021}).

\bibitem{Montgomery2021}
\bibinfo{author}{Montgomery, S.~M.} \emph{et~al.}
\newblock \bibinfo{journal}{\bibinfo{title}{{Magneto-Mechanical Metamaterials
  with Widely Tunable Mechanical Properties and Acoustic Bandgaps}}}.
\newblock {\emph{Advanced Functional Materials}}
  \textbf{\bibinfo{volume}{31}}, \bibinfo{pages}{2005319},
  \url{https://doi.org/10.1002/adfm.202005319} (\bibinfo{year}{2021}).

\bibitem{Wu2021}
\bibinfo{author}{Wu, L.} \emph{et~al.}
\newblock \bibinfo{journal}{\bibinfo{title}{{A brief review of dynamic
  mechanical metamaterials for mechanical energy manipulation}}}.
\newblock {\emph{Materials Today}}
  \textbf{\bibinfo{volume}{44}}, \bibinfo{pages}{168--193},
  \url{https://doi.org/10.1016/j.mattod.2020.10.006} (\bibinfo{year}{2021}).

\bibitem{Yang2015}
\bibinfo{author}{Yang, D.} \emph{et~al.}
\newblock \bibinfo{journal}{\bibinfo{title}{{Buckling of Elastomeric Beams
  Enables Actuation of Soft Machines}}}.
\newblock {\emph{Advanced Materials}}
  \textbf{\bibinfo{volume}{27}}, \bibinfo{pages}{6323--6327},
  \url{https://doi.org/10.1002/adma.201503188} (\bibinfo{year}{2015}).

\bibitem{Terryn2017}
\bibinfo{author}{Terryn, S.}, \bibinfo{author}{Brancart, J.},
  \bibinfo{author}{Lefeber, D.}, \bibinfo{author}{{Van Assche}, G.} \&
  \bibinfo{author}{Vanderborght, B.}
\newblock \bibinfo{journal}{\bibinfo{title}{{Self-healing soft pneumatic
  robots}}}.
\newblock {\emph{Science Robotics}}
  \textbf{\bibinfo{volume}{2}}, \bibinfo{pages}{16},
  \url{https://doi.org/10.1126/scirobotics.aan4268} (\bibinfo{year}{2017}).

\bibitem{Kim2019}
\bibinfo{author}{Kim, Y.}, \bibinfo{author}{Parada, G.~A.},
  \bibinfo{author}{Liu, S.} \& \bibinfo{author}{Zhao, X.}
\newblock \bibinfo{journal}{\bibinfo{title}{{Ferromagnetic soft continuum
  robots}}}.
\newblock {\emph{Science Robotics}}
  \textbf{\bibinfo{volume}{4}}, \bibinfo{pages}{7329},
  \url{https://doi.org/10.1126/SCIROBOTICS.AAX7329} (\bibinfo{year}{2019}).

\bibitem{veerabagu2022}
\bibinfo{author}{Veerabagu, U.}, \bibinfo{author}{Palza, H.} \&
  \bibinfo{author}{Quero, F.}
\newblock \bibinfo{journal}{\bibinfo{title}{Auxetic polymer-based mechanical
  metamaterials for biomedical applications}}.
\newblock {\emph{ACS Biomaterials Science \& Engineering}}
  \textbf{\bibinfo{volume}{8}}, \bibinfo{pages}{2798--2824},
  \url{https://doi.org/10.1021/acsbiomaterials.2c00109} (\bibinfo{year}{2022}).

\bibitem{Babaee2013}
\bibinfo{author}{Babaee, S.} \emph{et~al.}
\newblock \bibinfo{journal}{\bibinfo{title}{{3D soft metamaterials with
  negative poisson's ratio}}}.
\newblock {\emph{Advanced Materials}}
  \textbf{\bibinfo{volume}{25}}, \bibinfo{pages}{5044--5049},
  \url{https://doi.org/10.1002/adma.201301986} (\bibinfo{year}{2013}).

\bibitem{Nicolaou2012}
\bibinfo{author}{Nicolaou, Z.~G.} \& \bibinfo{author}{Motter, A.~E.}
\newblock \bibinfo{journal}{\bibinfo{title}{{Mechanical metamaterials with
  negative compressibility transitions}}}.
\newblock {\emph{Nature Materials}}
  \textbf{\bibinfo{volume}{11}}, \bibinfo{pages}{608--613},
  \url{https://doi.org/10.1038/nmat3331} (\bibinfo{year}{2012}).
\newblock \eprint{1207.2185}.

\bibitem{Boyce2008}
\bibinfo{author}{Bertoldi, K.}, \bibinfo{author}{Boyce, M.~C.},
  \bibinfo{author}{Deschanel, S.}, \bibinfo{author}{Prange, S.~M.} \&
  \bibinfo{author}{Mullin, T.}
\newblock \bibinfo{journal}{\bibinfo{title}{{Mechanics of deformation-triggered
  pattern transformations and superelastic behavior in periodic elastomeric
  structures}}}.
\newblock {\emph{Journal of the Mechanics and Physics of
  Solids}} \textbf{\bibinfo{volume}{56}}, \bibinfo{pages}{2642--2668},
  \url{https://doi.org/10.1016/j.jmps.2008.03.006} (\bibinfo{year}{2008}).

\bibitem{Overvelde2014}
\bibinfo{author}{Overvelde, J.~T.} \& \bibinfo{author}{Bertoldi, K.}
\newblock \bibinfo{journal}{\bibinfo{title}{{Relating pore shape to the
  non-linear response of periodic elastomeric structures}}}.
\newblock {\emph{Journal of the Mechanics and Physics of
  Solids}} \textbf{\bibinfo{volume}{64}}, \bibinfo{pages}{351--366},
  \url{https://doi.org/10.1016/j.jmps.2013.11.014} (\bibinfo{year}{2014}).

\bibitem{Kim2018}
\bibinfo{author}{Kim, Y.}, \bibinfo{author}{Yuk, H.}, \bibinfo{author}{Zhao,
  R.}, \bibinfo{author}{Chester, S.~A.} \& \bibinfo{author}{Zhao, X.}
\newblock \bibinfo{journal}{\bibinfo{title}{{Printing ferromagnetic domains for
  untethered fast-transforming soft materials}}}.
\newblock {\emph{Nature}} \textbf{\bibinfo{volume}{558}},
  \bibinfo{pages}{274--279}, \url{https://doi.org/10.1038/s41586-018-0185-0}
  (\bibinfo{year}{2018}).

\bibitem{Shim2013}
\bibinfo{author}{Shim, J.} \emph{et~al.}
\newblock \bibinfo{journal}{\bibinfo{title}{{Harnessing instabilities for
  design of soft reconfigurable auxetic/chiral materials}}}.
\newblock {\emph{Soft Matter}} \textbf{\bibinfo{volume}{9}},
  \bibinfo{pages}{8198--8202}, \url{https://doi.org/10.1039/c3sm51148k}
  (\bibinfo{year}{2013}).

\bibitem{Azulay2023}
\bibinfo{author}{Azulay, R.}, \bibinfo{author}{Combescure, C.} \&
  \bibinfo{author}{Dirrenberger, J.}
\newblock \bibinfo{journal}{\bibinfo{title}{Instability-induced pattern
  generation in architectured materials—a review of methods}}.
\newblock {\emph{International Journal of Solids and
  Structures}} \textbf{\bibinfo{volume}{274}}, \bibinfo{pages}{112240},
  \url{https://doi.org/10.1016/j.ijsolstr.2023.112240} (\bibinfo{year}{2023}).

\bibitem{azulay2024predicting}
\bibinfo{author}{Azulay, R.} \& \bibinfo{author}{Combescure, C.}
\newblock \bibinfo{journal}{\bibinfo{title}{Predicting the post-bifurcated
  patterns of architectured materials using group-theoretic tools}}.
\newblock {\emph{Journal of the Mechanics and Physics of
  Solids}} \textbf{\bibinfo{volume}{187}}, \bibinfo{pages}{105631}
  (\bibinfo{year}{2024}).

\bibitem{wallpapergroups}
\bibinfo{author}{{International Union of Crystallography}}.
\newblock \bibinfo{title}{International tables for crystallography, index of
  the 17 plane groups}.
\newblock In \emph{\bibinfo{booktitle}{International Tables for
  Crystallography}}, chap. \bibinfo{chapter}{2.2}, \bibinfo{pages}{175--192},
  \url{https://doi.org/10.1107/97809553602060000927}
  (\bibinfo{publisher}{International Union of Crystallography},
  \bibinfo{year}{2016}).

\bibitem{schattschneider1978plane}
\bibinfo{author}{Schattschneider, D.}
\newblock \bibinfo{journal}{\bibinfo{title}{The plane symmetry groups: their
  recognition and notation}}.
\newblock {\emph{The American Mathematical Monthly}}
  \textbf{\bibinfo{volume}{85}}, \bibinfo{pages}{439--450}
  (\bibinfo{year}{1978}).

\bibitem{Jin2019}
\bibinfo{author}{Jin, S.}, \bibinfo{author}{Korkolis, Y.~P.} \&
  \bibinfo{author}{Li, Y.}
\newblock \bibinfo{journal}{\bibinfo{title}{{Shear resistance of an auxetic
  chiral mechanical metamaterial}}}.
\newblock {\emph{International Journal of Solids and
  Structures}} \textbf{\bibinfo{volume}{174-175}}, \bibinfo{pages}{28--37},
  \url{https://doi.org/10.1016/j.ijsolstr.2019.06.005} (\bibinfo{year}{2019}).

\bibitem{han2022lightweight}
\bibinfo{author}{Han, D.} \emph{et~al.}
\newblock \bibinfo{journal}{\bibinfo{title}{Lightweight auxetic metamaterials:
  Design and characteristic study}}.
\newblock {\emph{Composite Structures}}
  \textbf{\bibinfo{volume}{293}}, \bibinfo{pages}{115706}
  (\bibinfo{year}{2022}).

\bibitem{kittel2018introduction}
\bibinfo{author}{Kittel, C.} \& \bibinfo{author}{McEuen, P.}
\newblock \emph{\bibinfo{title}{Introduction to solid state physics}}
  (\bibinfo{publisher}{John Wiley \& Sons}, \bibinfo{year}{2018}).

\bibitem{Maurin2018}
\bibinfo{author}{Maurin, F.}, \bibinfo{author}{Claeys, C.},
  \bibinfo{author}{Deckers, E.} \& \bibinfo{author}{Desmet, W.}
\newblock \bibinfo{journal}{\bibinfo{title}{{Probability that a band-gap
  extremum is located on the irreducible Brillouin-zone contour for the 17
  different plane crystallographic lattices}}}.
\newblock {\emph{International Journal of Solids and
  Structures}} \textbf{\bibinfo{volume}{135}}, \bibinfo{pages}{26--36},
  \url{https://doi.org/10.1016/j.ijsolstr.2017.11.006} (\bibinfo{year}{2018}).

\bibitem{Moore2024}
\bibinfo{author}{Moore, D.~B.}, \bibinfo{author}{Starkey, T.~A.} \&
  \bibinfo{author}{Chaplain, G.~J.}
\newblock \bibinfo{journal}{\bibinfo{title}{{Acoustic metasurfaces with Frieze
  symmetries}}}.
\newblock {\emph{The Journal of the Acoustical Society of
  America}} \textbf{\bibinfo{volume}{155}}, \bibinfo{pages}{568--574},
  \url{https://doi.org/10.1121/10.0024359} (\bibinfo{year}{2024}).

\bibitem{Liu2021a}
\bibinfo{author}{Liu, H.}, \bibinfo{author}{Plucinsky, P.},
  \bibinfo{author}{Feng, F.} \& \bibinfo{author}{James, R.~D.}
\newblock \bibinfo{journal}{\bibinfo{title}{{Origami and materials science}}}.
\newblock {\emph{Philosophical Transactions of the Royal Society
  A: Mathematical, Physical and Engineering Sciences}}
  \textbf{\bibinfo{volume}{379}}, \url{https://doi.org/10.1098/rsta.2020.0113}
  (\bibinfo{year}{2021}).
\newblock \eprint{2008.06026}.

\bibitem{Liu2024}
\bibinfo{author}{Liu, H.} \& \bibinfo{author}{James, R.~D.}
\newblock \bibinfo{journal}{\bibinfo{title}{{Design of origami structures with
  curved tiles between the creases}}}.
\newblock {\emph{Journal of the Mechanics and Physics of
  Solids}} \textbf{\bibinfo{volume}{185}}, \bibinfo{pages}{105559},
  \url{https://doi.org/10.1016/j.jmps.2024.105559} (\bibinfo{year}{2024}).

\bibitem{Wang2020b}
\bibinfo{author}{Wang, L.}, \bibinfo{author}{Chan, Y.~C.},
  \bibinfo{author}{Liu, Z.}, \bibinfo{author}{Zhu, P.} \&
  \bibinfo{author}{Chen, W.}
\newblock \bibinfo{journal}{\bibinfo{title}{{Data-driven metamaterial design
  with Laplace-Beltrami spectrum as “shape-DNA”}}}.
\newblock {\emph{Structural and Multidisciplinary Optimization}}
  \textbf{\bibinfo{volume}{61}}, \bibinfo{pages}{2613--2628},
  \url{https://doi.org/10.1007/s00158-020-02523-5} (\bibinfo{year}{2020}).

\bibitem{Chan2021}
\bibinfo{author}{Chan, Y.~C.}, \bibinfo{author}{Ahmed, F.},
  \bibinfo{author}{Wang, L.} \& \bibinfo{author}{Chen, W.}
\newblock \bibinfo{journal}{\bibinfo{title}{{METASET: Exploring shape and
  property spaces for data-driven metamaterials design}}}.
\newblock {\emph{Journal of Mechanical Design}}
  \textbf{\bibinfo{volume}{143}}, \url{https://doi.org/10.1115/1.4048629}
  (\bibinfo{year}{2021}).
\newblock \eprint{2006.02142}.

\bibitem{Bastek2023}
\bibinfo{author}{Bastek, J.~H.} \& \bibinfo{author}{Kochmann, D.~M.}
\newblock \bibinfo{journal}{\bibinfo{title}{{Inverse design of nonlinear
  mechanical metamaterials via video denoising diffusion models}}}.
\newblock {\emph{Nature Machine Intelligence}}
  \textbf{\bibinfo{volume}{5}}, \bibinfo{pages}{1466--1475},
  \url{https://doi.org/10.1038/s42256-023-00762-x} (\bibinfo{year}{2023}).
\newblock \eprint{2305.19836}.

\bibitem{Kollmann2020a}
\bibinfo{author}{Kollmann, H.~T.}, \bibinfo{author}{Abueidda, D.~W.},
  \bibinfo{author}{Koric, S.}, \bibinfo{author}{Guleryuz, E.} \&
  \bibinfo{author}{Sobh, N.~A.}
\newblock \bibinfo{journal}{\bibinfo{title}{{Deep learning for topology
  optimization of 2D metamaterials}}}.
\newblock {\emph{Materials and Design}}
  \textbf{\bibinfo{volume}{196}}, \bibinfo{pages}{109098},
  \url{https://doi.org/10.1016/j.matdes.2020.109098} (\bibinfo{year}{2020}).

\bibitem{Lyu123}
\bibinfo{author}{Lyu, X.} \& \bibinfo{author}{Ren, X.}
\newblock \bibinfo{journal}{\bibinfo{title}{{Microstructure reconstruction of
  2D/3D random materials via diffusion-based deep generative models}}}.
\newblock {\emph{Scientific Reports}}
  \textbf{\bibinfo{volume}{14}}, \bibinfo{pages}{5041},
  \url{https://doi.org/10.1038/s41598-024-54861-9} (\bibinfo{year}{2024}).
\newblock \eprint{2311.17319}.

\bibitem{Ohno2001}
\bibinfo{author}{Ohno, N.}, \bibinfo{author}{Okumura, D.} \&
  \bibinfo{author}{Noguchi, H.}
\newblock \bibinfo{journal}{\bibinfo{title}{Microscopic symmetric bifurcation
  condition of cellular solids based on a homogenization theory of finite
  deformation}}.
\newblock {\emph{Journal of the Mechanics and Physics of
  Solids}} \textbf{\bibinfo{volume}{50}}, \bibinfo{pages}{1125--1153},
  \url{https://doi.org/10.1016/S0022-5096(01)00106-5} (\bibinfo{year}{2002}).

\bibitem{vangelatos2020regulating}
\bibinfo{author}{Vangelatos, Z.}, \bibinfo{author}{Komvopoulos, K.} \&
  \bibinfo{author}{Grigoropoulos, C.}
\newblock \bibinfo{journal}{\bibinfo{title}{Regulating the mechanical behavior
  of metamaterial microlattices by tactical structure modification}}.
\newblock {\emph{Journal of the Mechanics and Physics of
  Solids}} \textbf{\bibinfo{volume}{144}}, \bibinfo{pages}{104112}
  (\bibinfo{year}{2020}).

\bibitem{tyburec2025modular}
\bibinfo{author}{Tyburec, M.}, \bibinfo{author}{Do{\v{s}}k{\'a}{\v{r}}, M.},
  \bibinfo{author}{Somr, M.}, \bibinfo{author}{Kru{\v{z}}{\'\i}k, M.} \&
  \bibinfo{author}{Zeman, J.}
\newblock \bibinfo{journal}{\bibinfo{title}{Modular-topology optimization for
  additive manufacturing of reusable mechanisms}}.
\newblock {\emph{Computers \& Structures}}
  \textbf{\bibinfo{volume}{307}}, \bibinfo{pages}{107630},
  \url{https://doi.org/10.1016/j.compstruc.2024.107630} (\bibinfo{year}{2025}).

\bibitem{Zheng2023}
\bibinfo{author}{Zheng, L.}, \bibinfo{author}{Karapiperis, K.},
  \bibinfo{author}{Kumar, S.} \& \bibinfo{author}{Kochmann, D.~M.}
\newblock \bibinfo{journal}{\bibinfo{title}{{Unifying the design space and
  optimizing linear and nonlinear truss metamaterials by generative
  modeling}}}.
\newblock {\emph{Nature Communications}}
  \textbf{\bibinfo{volume}{14}}, \bibinfo{pages}{1--14},
  \url{https://doi.org/10.1038/s41467-023-42068-x} (\bibinfo{year}{2023}).

\bibitem{Frankel2019}
\bibinfo{author}{Frankel, A.~L.}, \bibinfo{author}{Jones, R.~E.},
  \bibinfo{author}{Alleman, C.} \& \bibinfo{author}{Templeton, J.~A.}
\newblock \bibinfo{journal}{\bibinfo{title}{{Predicting the mechanical response
  of oligocrystals with deep learning}}}.
\newblock {\emph{Computational Materials Science}}
  \textbf{\bibinfo{volume}{169}},
  \url{https://doi.org/10.1016/j.commatsci.2019.109099} (\bibinfo{year}{2019}).
\newblock \eprint{1901.10669}.

\bibitem{Wilt2020a}
\bibinfo{author}{Wilt, J.~K.}, \bibinfo{author}{Yang, C.} \&
  \bibinfo{author}{Gu, G.~X.}
\newblock \bibinfo{journal}{\bibinfo{title}{{Accelerating Auxetic Metamaterial
  Design with Deep Learning}}}.
\newblock {\emph{Advanced Engineering Materials}}
  \textbf{\bibinfo{volume}{22}}, \bibinfo{pages}{1901266},
  \url{https://doi.org/10.1002/adem.201901266} (\bibinfo{year}{2020}).

\bibitem{Pfaff2020}
\bibinfo{author}{Pfaff, T.}, \bibinfo{author}{Fortunato, M.},
  \bibinfo{author}{Sanchez-Gonzalez, A.} \& \bibinfo{author}{Battaglia, P.}
\newblock \bibinfo{title}{Learning mesh-based simulation with graph networks}.
\newblock In \emph{\bibinfo{booktitle}{International conference on learning
  representations}} (\bibinfo{year}{2020}).

\bibitem{Vlassis2020}
\bibinfo{author}{Vlassis, N.~N.}, \bibinfo{author}{Ma, R.} \&
  \bibinfo{author}{Sun, W.}
\newblock \bibinfo{journal}{\bibinfo{title}{{Geometric deep learning for
  computational mechanics Part I: anisotropic hyperelasticity}}}.
\newblock {\emph{Computer Methods in Applied Mechanics and
  Engineering}} \textbf{\bibinfo{volume}{371}}, \bibinfo{pages}{113299},
  \url{https://doi.org/10.1016/j.cma.2020.113299} (\bibinfo{year}{2020}).
\newblock \eprint{2001.04292}.

\bibitem{Pandey2021}
\bibinfo{author}{Pandey, A.} \& \bibinfo{author}{Pokharel, R.}
\newblock \bibinfo{journal}{\bibinfo{title}{{Machine learning based surrogate
  modeling approach for mapping crystal deformation in three dimensions}}}.
\newblock {\emph{Scripta Materialia}}
  \textbf{\bibinfo{volume}{193}}, \bibinfo{pages}{1--5},
  \url{https://doi.org/10.1016/j.scriptamat.2020.10.028}
  (\bibinfo{year}{2021}).

\bibitem{Yang2021}
\bibinfo{author}{Yang, Z.}, \bibinfo{author}{Yu, C.~H.} \&
  \bibinfo{author}{Buehler, M.~J.}
\newblock \bibinfo{journal}{\bibinfo{title}{{Deep learning model to predict
  complex stress and strain fields in hierarchical composites}}}.
\newblock {\emph{Science Advances}}
  \textbf{\bibinfo{volume}{7}}, \url{https://doi.org/10.1126/SCIADV.ABD7416}
  (\bibinfo{year}{2021}).

\bibitem{Mianroodi2021}
\bibinfo{author}{Mianroodi, J.~R.}, \bibinfo{author}{{H. Siboni}, N.} \&
  \bibinfo{author}{Raabe, D.}
\newblock \bibinfo{journal}{\bibinfo{title}{{Teaching solid mechanics to
  artificial intelligence—a fast solver for heterogeneous materials}}}.
\newblock {\emph{npj Computational Materials}}
  \textbf{\bibinfo{volume}{7}},
  \url{https://doi.org/10.1038/s41524-021-00571-z} (\bibinfo{year}{2021}).
\newblock \eprint{2103.09147}.

\bibitem{Mianroodi2022}
\bibinfo{author}{Mianroodi, J.~R.}, \bibinfo{author}{Rezaei, S.},
  \bibinfo{author}{Siboni, N.~H.}, \bibinfo{author}{Xu, B.~X.} \&
  \bibinfo{author}{Raabe, D.}
\newblock \bibinfo{journal}{\bibinfo{title}{{Lossless multi-scale constitutive
  elastic relations with artificial intelligence}}}.
\newblock {\emph{npj Computational Materials}}
  \textbf{\bibinfo{volume}{8}}, \bibinfo{pages}{1--12},
  \url{https://doi.org/10.1038/s41524-022-00753-3} (\bibinfo{year}{2022}).
\newblock \eprint{2108.02837}.

\bibitem{Khorrami2023}
\bibinfo{author}{Khorrami, M.~S.} \emph{et~al.}
\newblock \bibinfo{journal}{\bibinfo{title}{{An artificial neural network for
  surrogate modeling of stress fields in viscoplastic polycrystalline
  materials}}}.
\newblock {\emph{npj Computational Materials}}
  \textbf{\bibinfo{volume}{9}}, \bibinfo{pages}{1--10},
  \url{https://doi.org/10.1038/s41524-023-00991-z} (\bibinfo{year}{2023}).
\newblock \eprint{2208.13490}.

\bibitem{Bastek2022}
\bibinfo{author}{Bastek, J.~H.}, \bibinfo{author}{Kumar, S.},
  \bibinfo{author}{Telgen, B.}, \bibinfo{author}{Glaesener, R.~N.} \&
  \bibinfo{author}{Kochmann, D.~M.}
\newblock \bibinfo{journal}{\bibinfo{title}{{Inverting the structure–property
  map of truss metamaterials by deep learning}}}.
\newblock {\emph{Proceedings of the National Academy of Sciences
  of the United States of America}} \textbf{\bibinfo{volume}{119}},
  \bibinfo{pages}{e2111505119}, \url{https://doi.org/10.1073/pnas.2111505119}
  (\bibinfo{year}{2022}).

\bibitem{Thomas2023}
\bibinfo{author}{Thomas, A.} \emph{et~al.}
\newblock \bibinfo{journal}{\bibinfo{title}{{Materials fatigue prediction using
  graph neural networks on microstructure representations}}}.
\newblock {\emph{Scientific Reports 2023 13:1}}
  \textbf{\bibinfo{volume}{13}}, \bibinfo{pages}{1--16},
  \url{https://doi.org/10.1038/s41598-023-39400-2} (\bibinfo{year}{2023}).

\bibitem{Karapiperis}
\bibinfo{author}{Karapiperis, K.} \& \bibinfo{author}{Kochmann, D.~M.}
\newblock \bibinfo{journal}{\bibinfo{title}{{Prediction and control of fracture
  paths in disordered architected materials using graph neural networks}}}.
\newblock {\emph{Communications Engineering}}
  \textbf{\bibinfo{volume}{2}},
  \url{https://doi.org/10.1038/s44172-023-00085-0} (\bibinfo{year}{2023}).

\bibitem{Hendriks2025}
\bibinfo{author}{Hendriks, F.}, \bibinfo{author}{Menkovski, V.},
  \bibinfo{author}{Do{\v{s}}k{\'{a}}ř, M.}, \bibinfo{author}{Geers, M.~G.} \&
  \bibinfo{author}{Roko{\v{s}}, O.}
\newblock \bibinfo{journal}{\bibinfo{title}{{Similarity equivariant graph
  neural networks for homogenization of metamaterials}}}.
\newblock {\emph{Computer Methods in Applied Mechanics and
  Engineering}} \textbf{\bibinfo{volume}{439}}, \bibinfo{pages}{117867},
  \url{https://doi.org/10.1016/j.cma.2025.117867} (\bibinfo{year}{2025}).
\newblock \eprint{arXiv:2404.17365}.

\bibitem{Sigmund1994}
\bibinfo{author}{Sigmund, O.}
\newblock \bibinfo{journal}{\bibinfo{title}{{Materials with prescribed
  constitutive parameters: An inverse homogenization problem}}}.
\newblock {\emph{International Journal of Solids and
  Structures}} \textbf{\bibinfo{volume}{31}}, \bibinfo{pages}{2313--2329},
  \url{https://doi.org/10.1016/0020-7683(94)90154-6} (\bibinfo{year}{1994}).

\bibitem{Andreassen2014}
\bibinfo{author}{Andreassen, E.}, \bibinfo{author}{Lazarov, B.~S.} \&
  \bibinfo{author}{Sigmund, O.}
\newblock \bibinfo{journal}{\bibinfo{title}{{Design of manufacturable 3D
  extremal elastic microstructure}}}.
\newblock {\emph{Mechanics of Materials}}
  \textbf{\bibinfo{volume}{69}}, \bibinfo{pages}{1--10},
  \url{https://doi.org/10.1016/j.mechmat.2013.09.018} (\bibinfo{year}{2014}).

\bibitem{Chen2018d}
\bibinfo{author}{Chen, Q.}, \bibinfo{author}{Zhang, X.} \&
  \bibinfo{author}{Zhu, B.}
\newblock \bibinfo{journal}{\bibinfo{title}{{Design of buckling-induced
  mechanical metamaterials for energy absorption using topology
  optimization}}}.
\newblock {\emph{Structural and Multidisciplinary Optimization}}
  \textbf{\bibinfo{volume}{58}}, \bibinfo{pages}{1395--1410},
  \url{https://doi.org/10.1007/s00158-018-1970-y} (\bibinfo{year}{2018}).

\bibitem{Thomsen2018}
\bibinfo{author}{Thomsen, C.~R.}, \bibinfo{author}{Wang, F.} \&
  \bibinfo{author}{Sigmund, O.}
\newblock \bibinfo{journal}{\bibinfo{title}{{Buckling strength topology
  optimization of 2D periodic materials based on linearized bifurcation
  analysis}}}.
\newblock {\emph{Computer Methods in Applied Mechanics and
  Engineering}} \textbf{\bibinfo{volume}{339}}, \bibinfo{pages}{115--136},
  \url{https://doi.org/10.1016/j.cma.2018.04.031} (\bibinfo{year}{2018}).

\bibitem{Wang2021}
\bibinfo{author}{Wang, F.} \& \bibinfo{author}{Sigmund, O.}
\newblock \bibinfo{journal}{\bibinfo{title}{{3D architected isotropic materials
  with tunable stiffness and buckling strength}}}.
\newblock {\emph{Journal of the Mechanics and Physics of
  Solids}} \textbf{\bibinfo{volume}{152}}, \bibinfo{pages}{104415},
  \url{https://doi.org/10.1016/j.jmps.2021.104415} (\bibinfo{year}{2021}).

\bibitem{Xue2022b}
\bibinfo{author}{Xue, T.} \& \bibinfo{author}{Mao, S.}
\newblock \bibinfo{journal}{\bibinfo{title}{{Mapped shape optimization method
  for the rational design of cellular mechanical metamaterials under large
  deformation}}}.
\newblock {\emph{International Journal for Numerical Methods in
  Engineering}} \textbf{\bibinfo{volume}{123}}, \bibinfo{pages}{2357--2380},
  \url{https://doi.org/10.1002/nme.6941} (\bibinfo{year}{2022}).

\bibitem{datasetzenodo}
\bibinfo{author}{Hendriks, F.} \emph{et~al.}
\newblock \bibinfo{title}{Wallpaper group-based mechanical metamaterials and
  their mechanical responses}.
\newblock \bibinfo{howpublished}{\emph{Zenodo}
  \url{https://zenodo.org/records/15849549}}, (\bibinfo{year}{2025}).

\bibitem{Abu-Mualla2024}
\bibinfo{author}{Abu-Mualla, M.} \& \bibinfo{author}{Huang, J.}
\newblock \bibinfo{journal}{\bibinfo{title}{{A Dataset Generation Framework for
  Symmetry-Induced Mechanical Metamaterials}}}.
\newblock {\emph{Journal of Mechanical Design}}
  \bibinfo{pages}{1--24}, \url{https://doi.org/10.1115/1.4066169}
  (\bibinfo{year}{2024}).

\bibitem{KarimiMahabadi2025}
\bibinfo{author}{{Karimi Mahabadi}, R.} \emph{et~al.}
\newblock \bibinfo{journal}{\bibinfo{title}{{Graph-based design of irregular
  metamaterials}}}.
\newblock {\emph{International Journal of Mechanical Sciences}}
  \textbf{\bibinfo{volume}{295}}, \bibinfo{pages}{110203},
  \url{https://doi.org/10.1016/j.ijmecsci.2025.110203} (\bibinfo{year}{2025}).

\bibitem{cgal:c-sspo2-24b}
\bibinfo{author}{Cacciola, F.}, \bibinfo{author}{Loriot, S.} \&
  \bibinfo{author}{Rouxel-Labb{\'e}, M.}
\newblock \bibinfo{title}{{2D} straight skeleton and polygon offsetting}.
\newblock In \emph{\bibinfo{booktitle}{{CGAL} User and Reference Manual}}
  (\bibinfo{publisher}{{CGAL Editorial Board}}, \bibinfo{year}{2024}),
  \bibinfo{edition}{{6.0.1}} edn.

\bibitem{Geuzaine2009}
\bibinfo{author}{Geuzaine, C.} \& \bibinfo{author}{Remacle, J.~F.}
\newblock \bibinfo{journal}{\bibinfo{title}{{Gmsh: A 3-D finite element mesh
  generator with built-in pre- and post-processing facilities}}}.
\newblock {\emph{International Journal for Numerical Methods in
  Engineering}} \textbf{\bibinfo{volume}{79}}, \bibinfo{pages}{1309--1331},
  \url{https://doi.org/10.1002/nme.2579} (\bibinfo{year}{2009}).

\bibitem{Polukhov2018}
\bibinfo{author}{Polukhov, E.}, \bibinfo{author}{Vallicotti, D.} \&
  \bibinfo{author}{Keip, M.~A.}
\newblock \bibinfo{journal}{\bibinfo{title}{{Computational stability analysis
  of periodic electroactive polymer composites across scales}}}.
\newblock {\emph{Computer Methods in Applied Mechanics and
  Engineering}} \textbf{\bibinfo{volume}{337}}, \bibinfo{pages}{165--197},
  \url{https://doi.org/10.1016/j.cma.2018.01.020} (\bibinfo{year}{2018}).

\bibitem{Geymonat1993a}
\bibinfo{author}{Geymonat, G.}, \bibinfo{author}{M{\"{u}}ller, S.} \&
  \bibinfo{author}{Triantafyllidis, N.}
\newblock \bibinfo{journal}{\bibinfo{title}{{Homogenization of nonlinearly
  elastic materials, microscopic bifurcation and macroscopic loss of rank-one
  convexity}}}.
\newblock {\emph{Archive for Rational Mechanics and Analysis}}
  \textbf{\bibinfo{volume}{122}}, \bibinfo{pages}{231--290},
  \url{https://doi.org/10.1007/BF00380256} (\bibinfo{year}{1993}).

\bibitem{Triantafyllidis2006}
\bibinfo{author}{Triantafyllidis, N.}, \bibinfo{author}{Nestorovi{\'{c}},
  M.~D.} \& \bibinfo{author}{Schraad, M.~W.}
\newblock \bibinfo{journal}{\bibinfo{title}{{Failure surfaces for finitely
  strained two-phase periodic solids under general in-plane loading}}}.
\newblock {\emph{Journal of Applied Mechanics, Transactions
  ASME}} \textbf{\bibinfo{volume}{73}}, \bibinfo{pages}{505--515},
  \url{https://doi.org/10.1115/1.2126695} (\bibinfo{year}{2006}).

\bibitem{Zhang2021a}
\bibinfo{author}{Zhang, G.}, \bibinfo{author}{Feng, N.} \&
  \bibinfo{author}{Khandelwal, K.}
\newblock \bibinfo{journal}{\bibinfo{title}{{A computational framework for
  homogenization and multiscale stability analyses of nonlinear periodic
  materials}}}.
\newblock {\emph{International Journal for Numerical Methods in
  Engineering}} \textbf{\bibinfo{volume}{122}}, \bibinfo{pages}{6527--6575},
  \url{https://doi.org/10.1002/nme.6802} (\bibinfo{year}{2021}).

\bibitem{Geers2010}
\bibinfo{author}{Geers, M. G.~D.}, \bibinfo{author}{Kouznetsova, V.~G.} \&
  \bibinfo{author}{Brekelmans, W. A.~M.}
\newblock \bibinfo{journal}{\bibinfo{title}{{Multi-scale computational
  homogenization: Trends and challenges}}}.
\newblock {\emph{Journal of Computational and Applied
  Mathematics}} \textbf{\bibinfo{volume}{234}}, \bibinfo{pages}{2175--2182},
  \url{https://doi.org/10.1016/j.cam.2009.08.077} (\bibinfo{year}{2010}).

\bibitem{wriggers2008nonlinear}
\bibinfo{author}{Wriggers, P.}
\newblock \emph{\bibinfo{title}{Nonlinear finite element methods}}
  (\bibinfo{publisher}{Springer Science \& Business Media},
  \bibinfo{year}{2008}).

\bibitem{Tadmor2012}
\bibinfo{author}{Tadmor, E.~B.}, \bibinfo{author}{Miller, R.~E.} \&
  \bibinfo{author}{Elliott, R.~S.}
\newblock \emph{\bibinfo{title}{{Continuum Mechanics and Thermodynamics}}}
  (\bibinfo{publisher}{Cambridge University Press}, \bibinfo{year}{2012}).

\bibitem{Kouznetsova2002}
\bibinfo{author}{Kouznetsova, V.~G.}
\newblock \emph{\bibinfo{title}{{Computational homogenization for the
  multi-scale analysis of multi-phase materials}}}.
\newblock Ph.D. thesis, \bibinfo{school}{Eindhoven University of Technology}
  (\bibinfo{year}{2002}).
\newblock \url{https://doi.org/10.6100/IR560009}.

\bibitem{Miehe2003}
\bibinfo{author}{Miehe, C.}
\newblock \bibinfo{journal}{\bibinfo{title}{{Computational micro-to-macro
  transitions for discretized micro-structures of heterogeneous materials at
  finite strains based on the minimization of averaged incremental energy}}}.
\newblock {\emph{Computer Methods in Applied Mechanics and
  Engineering}} \textbf{\bibinfo{volume}{192}}, \bibinfo{pages}{559--591},
  \url{https://doi.org/10.1016/S0045-7825(02)00564-9} (\bibinfo{year}{2003}).

\bibitem{Kunc2019}
\bibinfo{author}{Kunc, O.} \& \bibinfo{author}{Fritzen, F.}
\newblock \bibinfo{journal}{\bibinfo{title}{{Finite Strain Homogenization Using
  a Reduced Basis and Efficient Sampling}}}.
\newblock {\emph{Mathematical and Computational Applications}}
  \textbf{\bibinfo{volume}{24}}, \bibinfo{pages}{56},
  \url{https://doi.org/10.3390/mca24020056} (\bibinfo{year}{2019}).
\newblock \eprint{1904.01521}.

\bibitem{Rokos2020}
\bibinfo{author}{Roko{\v{s}}, O.}, \bibinfo{author}{Ameen, M.~M.},
  \bibinfo{author}{Peerlings, R. H.~J.} \& \bibinfo{author}{Geers, M. G.~D.}
\newblock \bibinfo{journal}{\bibinfo{title}{{Extended micromorphic
  computational homogenization for mechanical metamaterials exhibiting multiple
  geometric pattern transformations}}}.
\newblock {\emph{Extreme Mechanics Letters}}
  \textbf{\bibinfo{volume}{37}},
  \url{https://doi.org/10.1016/j.eml.2020.100708} (\bibinfo{year}{2020}).
\newblock \eprint{2004.05226}.

\bibitem{gorodtsov2019extreme}
\bibinfo{author}{Gorodtsov, V.~A.} \& \bibinfo{author}{Lisovenko, D.~S.}
\newblock \bibinfo{journal}{\bibinfo{title}{Extreme values of young’s modulus
  and poisson’s ratio of hexagonal crystals}}.
\newblock {\emph{Mechanics of Materials}}
  \textbf{\bibinfo{volume}{134}}, \bibinfo{pages}{1--8} (\bibinfo{year}{2019}).

\end{thebibliography}
\end{document}